\documentclass[onecolumn,authoryear]{els-mrw} 

\usepackage{amsmath,amssymb,amsfonts,amsthm,makeidx,graphicx}
\usepackage{txfonts}
\usepackage{helvet}
\usepackage{chemist}
\usepackage[version=4]{mhchem}
\newcommand{\msun}{\ifmmode M_\odot \else $M_\odot$\fi}
\begin{document}

\chapter{Basic stellar observables}\label{chap1}

\author[1]{Laurent Mahy}%
%\author[2]{Second Author}%

%\author[1,2]{Third Author}%

\address[1]{\orgname{Royal Observatory of Belgium}, \orgdiv{Department of Astronomy and Astrophysics}, \orgaddress{Avenue Circulaire/Ringlaan 3, B-1180 Brussels, Belgium}}
%\address[2]{\orgname{Name of Institute}, \orgdiv{Division or Department}, \orgaddress{Address of Institute}}

\articletag{Chapter Article tagline: update of previous edition,, reprint..}

\maketitle

\begin{glossary}[Glossary]
%\term{Europe} the model is a coherent view of capital markets data that allows users to interact with the content in a consistent manner.

%\term{Primates} regardless of the source. Essentially, of sources. Properly deployed.
\term{Wolf-Rayet stars:} classical Wolf-Rayet stars are an evolved stage of O-type stars and are grouped in three main types: nitrogen-rich Wolf-Rayet stars, carbon-rich Wolf-Rayet stars, and oxygen-rich Wolf-Rayet stars. \\
\term{Luminous Blue Variables:} rare class of massive, evolved stars that are characterized by their extreme luminosity, instability, and dramatic spectral and photometric variability on time span of a decade, called S-Doradus cycle.\\
\term{Red Supergiants:} massive stars in evolved stage that are characterized by their huge size, cool surface temperatures, reddish appearance and that have exhausted the hydrogen in their core and are now burning heavier elements such as helium. \\
\term{Pulsars:} highly magnetized, rotating neutron star, that emits periodic pulses of radiation in the radio waves and other electromagnetic bands at rates of up to one thousand pulses per second. \\
\term{Magnetars:} type of rapidly rotating neutron star having an exceptionally strong magnetic field, much stronger than that of typical neutron stars or even regular pulsars.\\
\term{Metallicity:} Abundance of elements heavier than hydrogen and helium in an environment, it serves as an important indicator of the environment's age, star formation history, and chemical evolution.\\

\end{glossary}

\begin{glossary}[Nomenclature]
\begin{tabular}{@{}lp{34pc}@{}}
GRB & Gamma-ray Burst\\
AGN & Active Galactic Nuclei\\
CMB & Cosmic Microwave Background \\
PNCS & Planetary Nebulae Central Stars \\
HRD & Hertzsprung-Russell diagram \\
RSG & Red Supergiant \\
LBV & Luminous Blue Variable star\\
WR & Wolf-Rayet star\\
%CMB & Cosmic Microwave Background \\
\end{tabular}
\end{glossary}

\begin{abstract}[Abstract]
Physical properties of stars such as luminosity, surface temperature, distance, or mass are measured from observations. These physical properties are of paramount importance to understand how stars are born, live, and die in the universe near and far. This chapter discusses the basic concepts used by astronomers to derive key information about stars from the light they emit. We present through a pedagogical approach the methods required for determining stellar brightness (apparent and absolute magnitudes), surface temperature (via black-body radiation and spectral classification), and distance (using parallax and standard candles). We finally review techniques for estimating stellar mass and radius, including the use of binary star systems and stellar evolution models.
\end{abstract}

\begin{glossary}[Key points]
\begin{itemize}
    \item The electromagnetic spectrum reveals different stellar characteristics, from temperature to stellar surface composition.
    \item The Hertzsprung-Russell diagram visualizes the evolutionary status of stars, linking temperature and luminosity.
    \item Fundamental parameters like luminosity, temperature, radius, and mass are critical for understanding the nature of stars and their evolutionary path.
    \item Distance measurements through parallax and standard candles provide spatial context for interpreting stellar properties.
    \item Binary stars are crucial for astrophysics because they provide a direct way to measure stellar masses but binary interactions, such as mass and angular momentum transfers, can drastically alter the evolution of the stars.
\end{itemize}
\end{glossary}

\section{Introduction}
Stars are the fundamental building blocks of the galaxies. They are responsible for the production of most of the chemical elements that are found in the universe, and that constitute the basis of life as we know it. From the closest neighbors within our Galaxy to the farthest reaches of the observable universe, stars offer a wealth of information about the processes that shape the cosmos. To unlock the secrets held within these distant objects, astronomers rely on a set of basic stellar observables that can be directly measured or inferred from observations. 

This chapter is intended to be a pedagogical introduction to stellar observables. It delves into the basic concepts, and observing techniques that are daily used by astronomers for understanding the stellar physics, and that constitute the observational framework of Astronomy. In this chapter, we will introduce fundamental notions to classify the stars, such as the electromagnetic window, brightness, color, distances, or kinematics, and we will give a detailed description of the physical properties of stars such as mass, luminosity, or effective temperature. These ingredients are essential to understand the life cycle of stars from their birth in molecular clouds to their final fates as remnants like white dwarfs, neutron stars or black holes. 

As we embark on this exploration of stellar observables, we will uncover the methods and principles that allow astronomers to transform light from these distant sources to get the whole picture of their stellar physics. Through this chapter, we aim to put the first pieces of the astronomical puzzle in place to explore the complex and dynamic universe that surrounds us.

\section{Observational frameworks of astronomy}\label{chap1:sec1}
All we know about stars is derived from observations of their electromagnetic spectrum. Human eyes are designed to only see a narrow band of the electromagnetic spectrum, covering the 380-780 nm wavelength range. However, the electromagnetic spectrum covers a much longer wavelength domain and is divided into separate bands: radio waves, microwaves, infrared (IR), visible light (optical), ultraviolet (UV), X-rays, and $\gamma$-rays, from high to low wavelengths. Some of these bands are observable from Earth, others are blocked by the Earth's atmosphere, thus requiring space-based observations. The electromagnetic waves in each of these bands have different characteristics and allow the scientists to study different kinds of phenomena. For instance:

\begin{itemize}
    \item {\bf $\gamma$-rays} ($\leq$ 0.01~nm) are only observable from space and allow us to study some of the most energetic processes in the universe such as short- and long-duration GRBs, pulsars, magnetars, AGNs, supernova remnants, etc.
    \item {\bf X-rays} (0.01 to 10 nm) are only observable from space and are used to detect high-energy photons coming from energetic phenomena in the universe such as accreting black holes or stars, colliding wind binaries, and from all the processes with hot highly ionized gas,
    \item {\bf Ultraviolet} (10 nm to 380 nm) is only observable from space and allows us detailed observations of hot star winds, white dwarfs, etc. 
    \item {\bf Optical} observations (380 to 780 nm) are useful to characterize the stellar photospheres and their properties, 
    \item {\bf Infrared} observations (780 nm to 1 mm) are crucial for studying cooler stars, dust clouds, and regions of star formation that are opaque in visible light,
    \item {\bf Microwaves} (1 mm to several cm) is very useful to explore the cold, diffuse and often hidden aspects of the universe such as the Cosmic Microwave Background (CMB) which is the afterglow of the Big Bang and represents one of the oldest signals we can detect; Molecular clouds, source of new generations of stars; Planets and Moons, among others,
    \item {\bf Radio waves} (several cm to several m) penetrate Earth's atmosphere with minimal absorption, allowing us for detailed observations of Interstellar medium (ISM), molecular clouds, synchrotron emission in the stellar winds, etc. 
\end{itemize}

\subsection{Electromagnetic windows and stellar spectrum}\label{sec1:subsec1}

\begin{figure}[b]
\centering
\includegraphics[width=15cm,trim=10 10 0 15,clip]{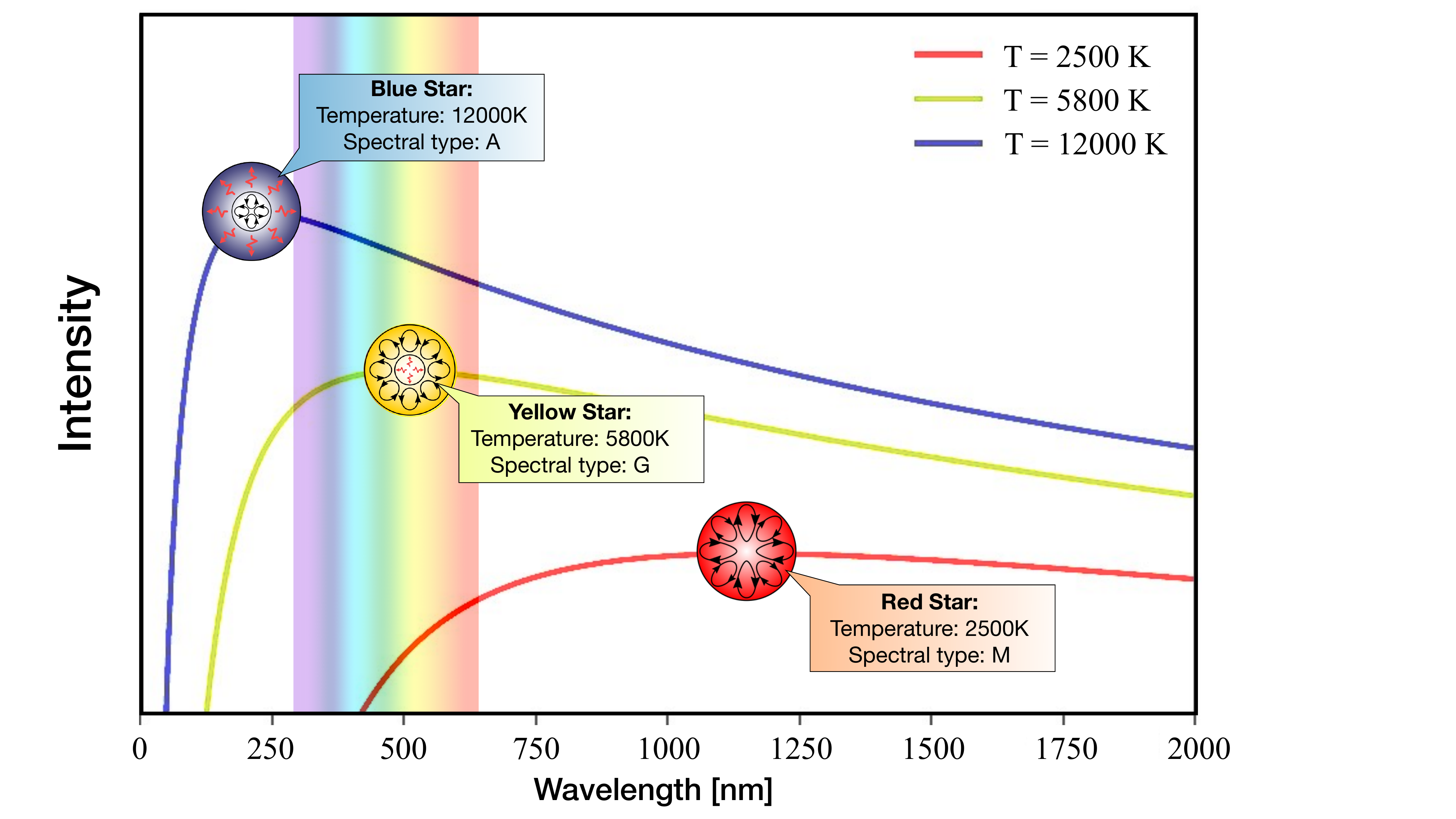}
\caption{Spectral distribution of the intensities of three black-body radiations computed from different black-body temperatures.}
\label{chap1:fig1}
\end{figure}

A star is a celestial body made of gas/plasma that is in hydrostatic equilibrium, i.e., its pressure-gradient force counterbalances its own gravity. Its density drops more or less smoothly from the center outwards, approaching the interstellar density far out from the star. Its radius is defined by its photosphere. It is from this region, that the continuum of the star is emitted. This continuum is approximately described by black-body radiation, following Planck's radiation law:
\begin{equation}\label{eqplanck}
    B(\lambda, T) = \frac{2~hc^2}{\lambda^5} \frac{1}{e^{hc/(\lambda~k_B~T)}-1}, 
\end{equation}
where $B$ is the black-body radiance expressed as a function of the wavelength, $h = 6.626 \times 10^{-34}$~m$^2$~kg~s$^{-1}$ is the Planck constant, $c = 299792.458$~km~s$^{-1}$ is the speed of light, $T$ is the temperature of the black body, and $k_B = 1.381 \times 10^{-23}$~m$^2$~kg~s$^{-2}$~K$^{-1}$ is the Boltzmann constant. The units of the Planck function are erg~s$^{-1}$~cm$^{-2}$~\AA$^{-1}$~steradian$^{-1}$.\\

A black body is an ideal object that absorbs all radiation falling upon it, in all frequencies, and that is able to re-emit the same amount of energy it absorbs with the same spectral shape. If we plot the amount of energy radiated by a black body as a function of wavelength (or frequency), we get a characteristic distribution that solely depends on the effective temperature of the object.

The law of Stefan-Boltzmann describes the relation between the energy emitted and the temperature of a black body. This law tells us that the energy flux of a black body is directly proportional to its temperature to the fourth power: 
\begin{equation}
E = \sigma T^4,
\end{equation}
where $T$ is the temperature of the object in Kelvin scale, $E$ is the energy flux and $\sigma = 5.669~10^{-8}$~W~m$^{-2}$~K$^{-4}$ is known as the Stefan-Boltzmann constant. It characterizes the total energy that is emitted per unit of time per unit of area. Through Equation~\ref{eqplanck}, one sees that hot massive stars peak in the blue wavelength range (UV), while cooler stars like red supergiants peak in the infrared (Fig.~\ref{chap1:fig1}). \\

\begin{figure}[b]
\centering
\includegraphics[width=15cm,trim=20 45 20 10,clip]{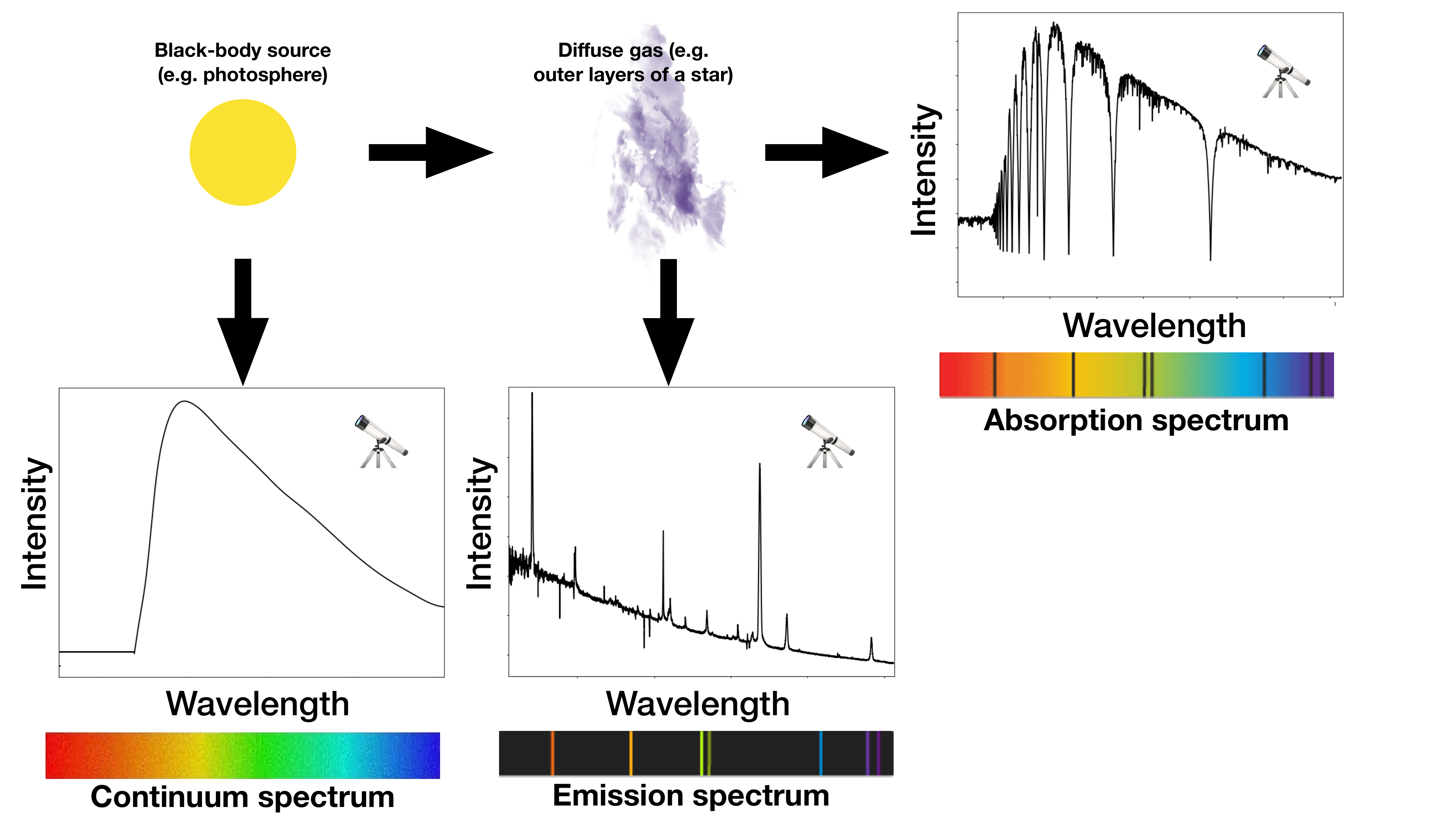}
\caption{The various types of stellar spectra and how they are produced from a same source.}
\label{chap1:fig2}
\end{figure}

In reality, stars are not perfect black bodies, so their emitted spectra do not just depend on the surface temperature, but contain detailed signatures of their physical properties such as their surface composition. Spectra of astronomical objects thus show continua described by the Stefan-Boltzmann law, and on top by the so-called spectral lines. A hot tenuous gas gives off an emission line spectrum, while a continuous spectrum that passes through a relatively cool gas results in an absorption spectrum superimposed on the continuous spectrum (Fig.~\ref{chap1:fig2}). Studying these stellar spectra allows us to determine the stellar composition, the physical properties of the stars as well as their kinematics.

\subsubsection{Absorption lines}\label{sec1:subsubsec1}
The spectrum of a star is thus composed of a continuum and of spectral lines. A majority of stars have stellar spectra exhibiting  absorption features. These lines are formed when photons with specific energies are absorbed by atoms or ions in the stellar atmosphere. These discrete energy transitions cause electrons to move from lower energy levels to higher ones. These are described as {\it bound-bound}  transitions. Excitation to a higher energy level can be provoked either by the absorption of kinetic energy ({\it collisional excitation}) or by the absorption of a photon ({\it radiative excitation}). Each element has a unique set of energy levels. Since a star’s spectrum emerges from its photosphere, its effective temperature will play a pivotal role in determining which atomic lines are present in the spectrum. So the pattern of absorption lines acts as a fingerprint, that characterizes the star. The strengths of these lines depend on the abundance of the element, on the surface temperature, and also on the electronic pressure. \\

Each spectral line of the elements appears at a well-determined wavelength or {\it rest} wavelength (measured in laboratories). Their shifts in position with respect to their rest wavelength, the so-called Doppler shift, describe the radial motions (or velocities) of the stars. When the spectral lines are blue-shifted, the star is moving towards the observer, while a red-shift of the spectral lines indicates that the star is moving away from the observer. This is particularly important to notably detect binary systems as explained in Section~\ref{chap1:sec4} or to a lesser extent pulsations. 

\subsubsection{Emission lines}\label{sec1:subsubsec2}
The spectra of certain types of stars can also exhibit emission lines. This emission spectrum is produced when an observer is not looking directly at the black body source but instead is diffused by a cloud of gas (Fig.~\ref{chap1:fig2}). Similarly to absorption lines, the emission lines are produced by de-excitation, either caused by a collision ({\it collisional de-excitation}) or by emission of a photon ({\it radiative de-excitation}). In this case, the electrons lose energy and drop down to lower energy levels. When an electron drops down between levels, it emits photon with the same amount of energy that it would need to absorb in order to move up between those same levels. Emission line spectra are characteristics of (1) evolved massive stars that have lost much of their hydrogen envelope and are in the later stages of stellar evolution like Wolf-Rayet (WR) stars, or Luminous Blue variable (LBV) stars, (2) T-Tauri stars that are young, pre-main-sequence stars that are still contracting and accreting material from their surrounding molecular clouds, (3) Planetary Nebulae Central Stars (PNCS), which are remnants of stars like the Sun at the center of planetary nebulae in the final stages of their evolution.

\subsubsection{Type of spectra}\label{sec1:subsubsec3}
Stellar spectra are thus a unique fingerprint to characterize and classify the stars. They mostly differ from each other with respect to the surface temperatures of the stars. From their optical spectra, astronomers has thus classified them according to a unique scheme. Knowing the spectral classification (or type) of a star tells us a lot about what kind of star we are dealing with such as its temperature range or if the star is big or small. The concept of a spectral type is to classify stars into groups according to the strengths of their spectral lines \citep{gray05}. 

\begin{figure}[htpb]
\centering
\includegraphics[width=16cm,trim=0 0 0 10,clip]{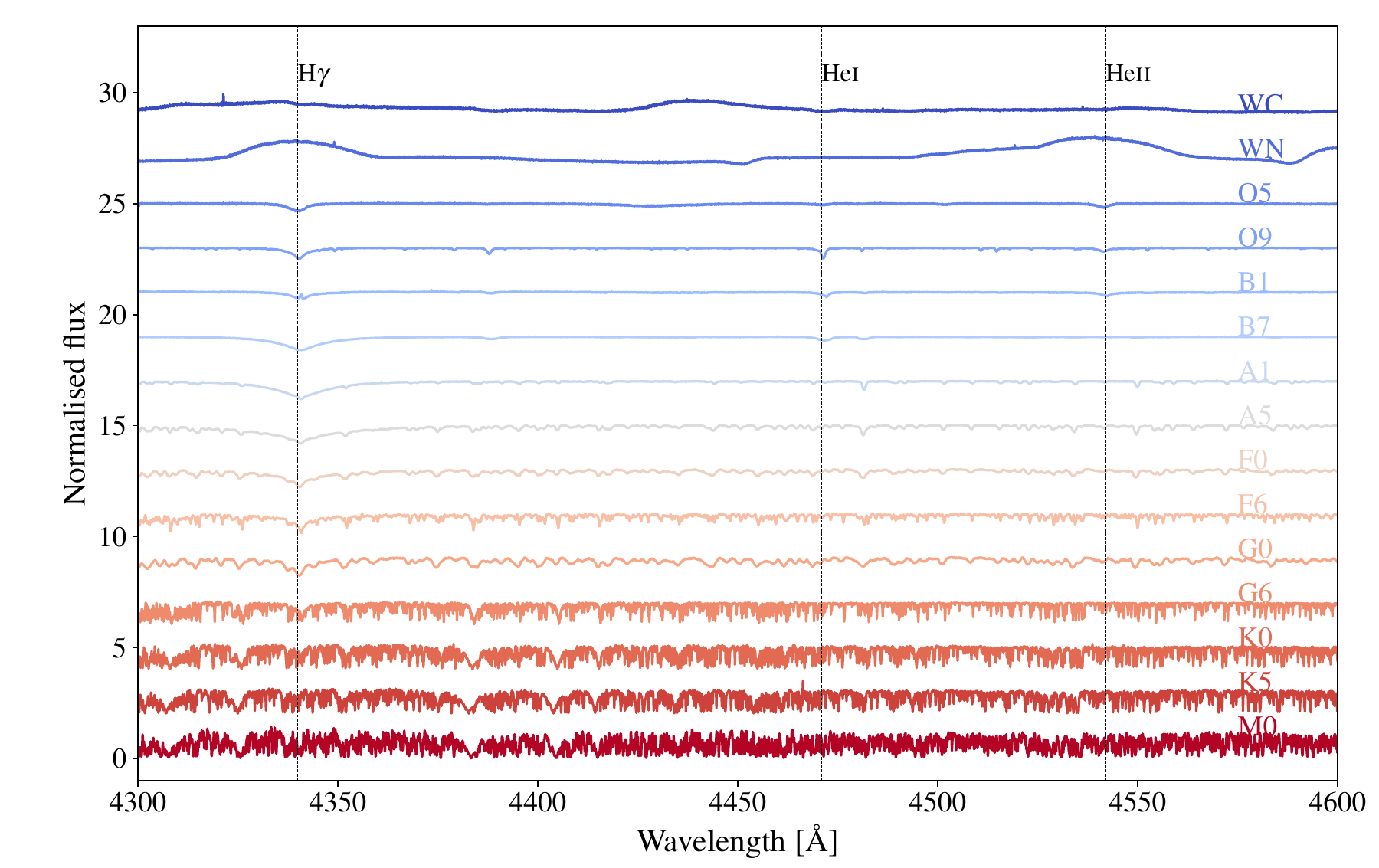}
\caption{Spectroscopic atlas in the optical range covering $4300 - 4600~\AA$. Spectra are from the MELCHIOR database \citep{royer24}. They are vertically shifted for clarity, and color-coded with respect to the surface temperature, blue color meaning hotter stars and red color cooler stars.}
\label{chap1:fig3}
\end{figure}
The first classification attempt was done, based on the strength of the hydrogen lines. Hydrogen ionizes at about 10~000~K. The stars with the strongest hydrogen lines were classified as A-type stars, while a M-type classification was attributed to the weakest hydrogen lines.

%When the temperature raises, fewer atoms have bound electrons, and the lines disappear from the spectra. 
The current temperature classification sequence, from the hottest to the coolest stars, follows the Morgan-Keeman system classification that uses the letters:
$$ \rm{ O \qquad B  \qquad A  \qquad  F \qquad G \qquad K \qquad M }\footnote{A useful mnemonic to remember the order of the spectral classes is "{\it Oh Be A Fine Girl/Guy, Kiss Me}".}.$$  
 
Each star in these groups has specific observational features that are summarized in Table~\ref{chap1:tab1}. Furthermore, spectral classification of very cool stars includes extensions to classes L (2,200--1,300~K) and T (1,300--800~K). Three other spectral types R, N and S, representing a side branch to the normal classification scheme, and not associated directly with temperature are also used. The R-type stars constitute a parallel branch to the K-type stars but showing carbon molecular bands, the N- and S-type stars represent two parallel groups to the M-type stars but showing, for one, cooler carbon molecular bands and, for the other, zirconium oxide bands. 

M-type stars have initial masses of less than 0.5~\msun, and are fully convective, i.e., they are stars in which convection occurs throughout their entire interior. They represent about 75\% of stars in the Milky Way.  Stars like our Sun have an initial mass between 0.5 and 1.5~\msun. They are classified as F-, G- or K-type stars and have a radiative core and a convective envelope. The stars more massive than 1.5~\msun\ are classified as A-, B- or O-type stars, and have a radiative envelope and a convective core (see Fig.~\ref{chap1:fig1}). O- and B-type stars represent only 0.15\% of the global population of stars in the Milky Way. Table~\ref{chap1:tab1} provides the standard Harvard spectral classification scheme for stars.

Among each of these temperature (or spectral) classifications, stellar spectra can be very different (Fig.~\ref{chap1:fig3}). It was therefore decided to divide each type into multiple spectral sub-types, noted by adding an Arabic numeral so that the sequence is decreasing with respect to the effective temperature. About 10 subdivisions per letter group are given but some sub-classes are rarely used. For example, the main sub-types for the cooler G-type stars are G0, G2, G5, G8, while for the K- and M-type stars, it goes from K0 or M0, to K5 or M5. In a few cases, half intervals (or more) are given full weight as a sub-class. Among massive stars for instance, the O-type stars are subdivided from O2, O3, O3.5, ..., to O9, O9.2, O9.5, and O9.7, while a sequence from B0, B0.2, B0.5, B0.7, B1, B1.5, ..., B8, B9 defines the B-type spectral classification. Those spectra toward the cool end of the (sub-)sequence are referred to as “late-type” spectra, and those on the hot end as “early-type” spectra. This designation is valid inside the global spectral classification but also inside the sub-types sequence. These terms came about when astronomers erroneously thought that stars began their lives as hot stars and cooled down during their lifespan.

To provide the complete spectral classification, the sequence consisting of a capital letter followed by an Arabic number must be followed by a Roman numeral that represents the luminosity classification. This sequence follows: 0, I, Ia, Ib, II, III, IV, V, VI, and VII in order of decreasing luminosity. 0 representing the hypergiants, I (a or b, according to their luminosity) the supergiants, II the bright giants, III the giants, IV the subgiants, and V the dwarfs, VI the sub-dwarfs, and VII the white dwarfs.  By definition, the Sun’s spectral type is G2~V. Often small capitals are added to the luminosity classes, based on the appearance of specific characteristics of the spectral lines. This way stars of which Balmer lines are observed in emission are labeled with small capital “e” behind the luminosity class, and hot stars with N~\textsc{iii} and He~\textsc{ii} lines in emission get an “f” , etc \citep[see][and references therein for a complete description of these suffixes]{vanderhucht96}. We refer to the chapter dedicated to "spectral classification of stars" in this book for more details. 

%%% table with spectral classification - temperature range + typical spectrum of each star %%%
{\scriptsize\begin{table}[htbp]
\TBL{\caption{The standard Harvard spectral classification scheme}\label{chap1:tab1}}
{\begin{tabular*}{\textwidth}{@{\extracolsep{\fill}}@{}llllll@{}}
\toprule
\multicolumn{1}{@{}l}{\TCH{Spec.}} &
\multicolumn{1}{l}{\TCH{$T_{\rm eff}$}} &
\multicolumn{1}{l}{\TCH{Mass}} &
\multicolumn{1}{l}{\TCH{Ages}} &
\multicolumn{1}{l}{\TCH{Spectral features}}&
\multicolumn{1}{l}{\TCH{Predominant lines}}\\
\colrule
O & $    \geq 30$~kK & $\geq 16~\msun$ & $1-10$~Myrs & Ionized helium & Ratio between He~{\sc{i}}~4471/He~{\sc{ii}}~4541 \\
& & & & &  decrease towards earlier types \\
B & $10 - 30$~kK & $2.1 - 16~\msun$ & $11-400$~Myrs  & Neutral helium, & Ratio between Si~{\sc{iv}}~4089, Si~{\sc{iii}}~4552,\\
& & & & hydrogen & Si~{\sc{ii}}~4128--4130 decrease with types \\
& & & & & He~{\sc{i}} lines reach maximum at B2\\
A & $ 7.5 - 10$~kK & $1.4-2.1~\msun$ & $0.4 - 3.0$~Gyrs  & Hydrogen & Ca~{\sc{ii}} K line present \\
& & & & & Neutral metals become stronger\\
F & $ 6 - 7.5$~kK & $1.04-1.4~\msun$ & $3 - 7$~Gyrs  & Neutral hydrogen, & G band (4305\AA) visible\\
& & & &  ionized calcium & Neutral metals increase with type \\
G & $ 5.2 -  6$~kK & $0.8-1.04~\msun$ & $7 - 15$~Gyrs  & Neutral hydrogen, &  Neutral metals increase with type \\
& & & & strong ionized calcium & \\
K & $ 3.7 -  5.2$~kK & $0.45-0.8~\msun$ & $\sim 17$~Gyrs  & Neutral metals,  & Ca~{\sc{ii}}~4226 increases with types\\
& & & & ionized calcium & \ce{TiO} starts near K5 in giants\\
M & $ \leq 3.7$~kK & $0.08-0.45~\msun$ & $\sim 56$~Gyrs  & Molecules and neutral metals & Ca~{\sc{ii}}~4226 increases rapidly with type \\
  &                &                   &                 &                               & \ce{TiO} starts near M0 in dwarfs, K5 in giants\\
\botrule
\end{tabular*}}
{
%\begin{tablenotes}
%\footnotetext[a]{Table footnote text...}
%\footnotetext{\source{Table source text...}}
%\end{tablenotes}
}
\end{table}}

%%%% Cite Gray 2005

\subsection{Magnitudes and different systems}\label{sec1:subsec2}
Scientists like to classify objects. The Greek astronomer Hipparchus ($\sim -100$~BC) grouped the visible stars into six classes based on their apparent brightnesses (from 1 for the brightest objects to 6 for the faintest ones). These classes are now known as classes of {\it apparent magnitudes}. The same principle has been kept nowadays to classify the stars: a first magnitude star is brighter by a factor $5 \times 10^{-0.4}$ than a sixth magnitude star. 

The magnitude is defined as a logarithmic scale representing the radiative energy coming from an astronomical source. It might be different depending on the wavelength range (or frequency range) that is considered. When we consider two sources, the difference in magnitude between source 2 and source 1 is given by $$m_2 - m_1 = -2.5 \log~\left(\frac{f_2}{f_1}\right),$$
where $f$ is the amount of radiative energy received per unit of time. Therefore, if two times more energy is received from source 2 than from source 1, the magnitude of source 2 is 0.75 mag (the unit is mag) lower than that of source 1.

When a magnitude at only one wavelength $\lambda$ is considered, one speaks about {\it monochromatic magnitude} $m_{\lambda}$. In practice what we observe are magnitudes related to a selected wavelength range. However, the apparent magnitudes in a specific band $X$, $m_X$, tell us nothing about the intrinsic brightness of the stars because, in addition of being dependent on the intrinsic brightness of the star, they also depend on the distance of the star. The main example is our Sun. Our star is not particularly bright in the Milky Way, but because its distance is extremely close to us, it appears spectacularly bright to our eyes. 

With that respect, the {\it absolute magnitude} $M$ of a star is defined as the magnitude one would measure if the star was located at a distance of 10~pc from us: 
\begin{eqnarray*}
    m - M &=& 5 \log_{10} \left(\frac{d}{10}\right),\\
    m - M &=& 5 \log_{10} \left( d \right) - 5,
\end{eqnarray*} 
where $m$ is the apparent magnitude, $M$ the absolute magnitude, and $d$ the distance of the star expressed in parsec\footnote{One parsec is $3.262$ light years, which corresponds to $3.085 \times 10^{13}$~km \citep{Seidelmann92}.}. The quantity $m-M$ is also called the distance modulus.\\
This can be also written 
\begin{equation}
    M = m - 5 \log_{10} \left( d \right) + 5.\label{eq2}
\end{equation} 

The use of photo-electric detectors {allowed us} to determine stellar magnitudes in different wavelength ranges. For a long time, the zero point to calibrate the magnitude system was chosen to be the star Vega (i.e. $f_1$ is the flux of Vega in a given band and $m_1$ the magnitude of Vega is chosen to be equal to 0 in every pass-band). However, the choice of Vega as the calibrator is problematic for two reasons: 1) Vega does not have a flat spectral energy distribution (especially in the UV and the infrared), so it does not make much sense to force it to be flat, and 2) Vega may be a $\delta$-Scuti star, which varies in brightness. To circumvent that problem, one decided to calibrate the system using the absolute physical flux from Vega across Vega's spectral energy distribution. In this system, the zero-point source is a theoretical source defined such that the spectral energy distribution is constant through all the wavelengths. This new photometric system is the AB system (where AB stands for "absolute"). The most famous pass-bands have been defined by Johnson in the U (365 nm), B (440 nm), V (560 nm), R (700 nm) wavelength domains, as well as in the infrared with the I (900 nm), J (1250 nm), H (1650 nm), K (2250 nm), L (3600 nm), and M (5000 nm) filters. In the extreme case where we use all the wavelengths of the electromagnetic spectrum, the magnitude that is considered is the {\it bolometric magnitude} $M_{\rm bol}$.  The difference between the bolometric magnitude and the visual absolute magnitude is called the bolometric correction, $$BC_V = M_{\rm bol} - M_V.$$

Given that the objects in the sky have different energy distributions, we must define the term {\it color index}. This term has been introduced to define the difference between the magnitudes of the same star at different wavelengths. The color indices thus are a measure of an intrinsic characteristic of the star. The index $B-V$ , e.g., is a good measure for the effective temperature of an "intermediate-type" star (see Sect.~\ref{sec2:subsec3}). %Two commonly used color indices are $U-B$ and $B-V$.  %The difference in apparent magnitude is a quantity that is easily measured. %If we apply the relation~\ref{eq1} on two different magnitudes of the same star and subtract term by term, we find $$M_2-M_1 = m_2-m_1.$$

\subsection{Interstellar reddening and extinction}\label{sec1:subsec3}
For the stars that are far away from us, it is also important to consider the interstellar absorption. Although one thinks interstellar space is vacuum, it is in fact filled by gas and dust that absorbs and scatters part of the light that comes from the stars. This effect is called {\it interstellar extinction}.  It results in a systematic error of the apparent magnitude of the stars. The latter being fainter than they actually are $$V = V_0 + A_V $$ where $V_0$ is the intrinsic apparent magnitude of the star, represented as if there were no intervening material between the observer and the star, and $A_V$ is the extinction in the $V$ band.\\

If interstellar extinction is not taken into account, it can result into an overestimation of the distance of the star. This absorption is caused by the Rayleigh diffusion, that strongly affects the radiation at short wavelengths, as red light passes through gas and dust more easily than blue light. The consequence is that blue light coming from distant objects is strongly absorbed and scattered by dust, making the objects look redder than they actually are. The impact of the interstellar extinction is therefore stronger for the blue (hot) stars than for cooler stars. The reddening of starlight due to the interstellar extinction is known as {\it interstellar reddening}. The reddening of an object can be determined by measuring the color index, e.g., $(B-V)$, of the object and compare it to its intrinsic color $(B-V)_0$:
$$E(B-V) = (B-V)-(B-V)_0.$$
Interstellar reddening and extinction are therefore linked by the equation: 
\begin{equation}
    A_V = R_V \times E(B-V),
\end{equation}
where $R_V$ is the stand-alone parameter of relative visibility. Its standard value is generally taken to be 3.1, but can vary from 1 to 5, depending on the exact nature of the interstellar medium \citep{maiz18}.

\subsection{The Hertzsprung-Russell diagram}\label{sec2:subsec6}

\begin{figure}[b]
\centering
\includegraphics[width=16cm,trim=380 360 60 20,clip]{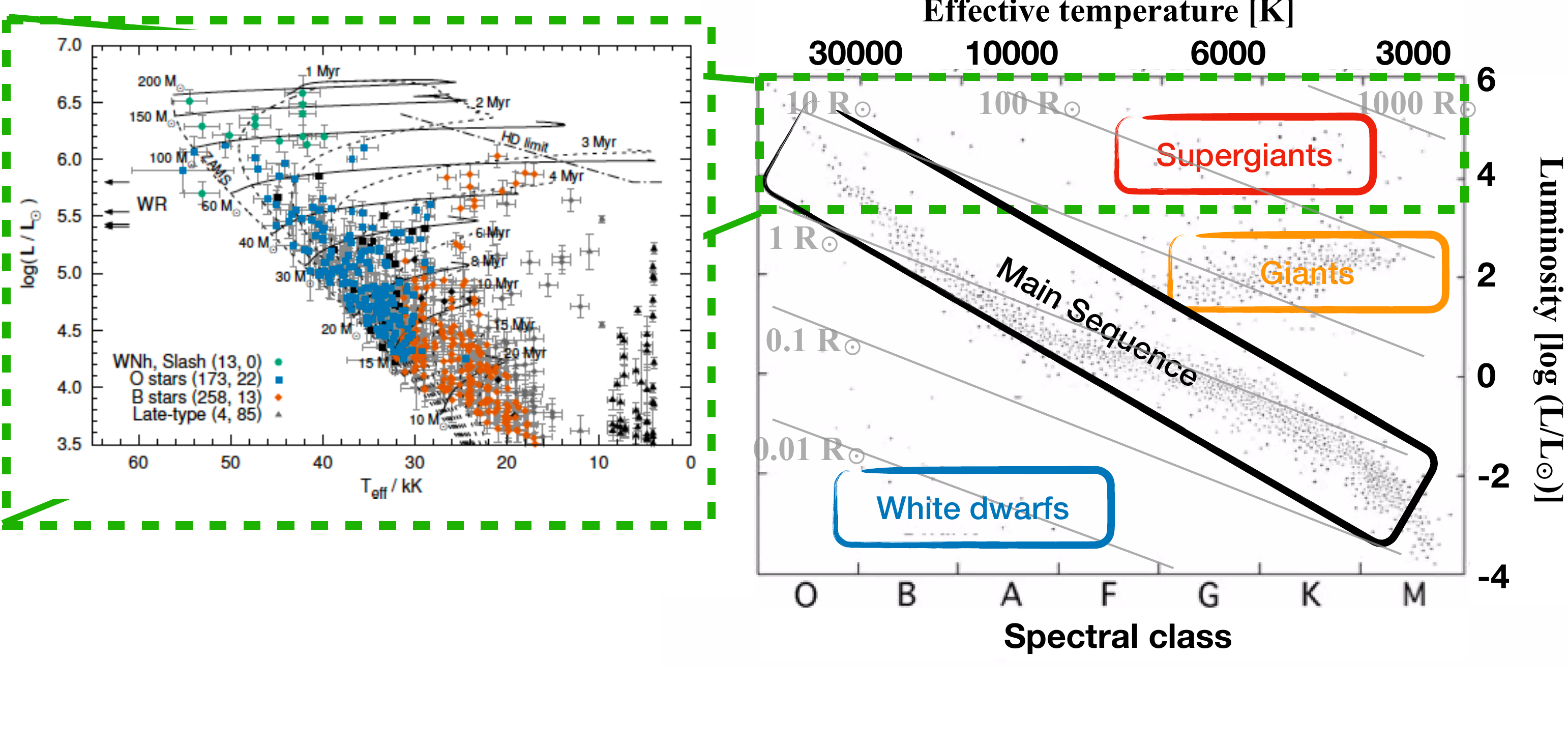}
\caption{An observational Hertzsprung–Russell diagram (right panel). Stars tend to fall only into certain regions of the diagram, marked in different colors. The most prominent is the diagonal in black, going from the upper-left (hot and bright) to the lower-right (cooler and less bright), called the Main Sequence. In the lower-left is where white dwarfs are found, and above the Main Sequence are the giants, and supergiants. The inset represents a zoom-in of the upper part of the HRD, where massive stars are located. It has been built from massive stars in the 30~Dor region in the LMC and is taken from \citet{schneider18}.}
\label{chap1:fig4}
\end{figure}

In their quest of classifying the stars, astronomers have created the so-called Hertzsprung-Russell diagram (HRD, named after the Danish astronomer Ejnar Hertzsprung and the American astronomer Henry Russell). This diagram represents an important diagnostic tool for the discussion of stellar evolution. The HRD (Fig.~\ref{chap1:fig4}) is built with an effective temperature range decreasing from left to right in abscissa, and the stellar luminosity in logarithmic scale in ordinates. Often the colour index $B-V$ (or $G_{BP} - G_{RP}$ if one uses the Gaia bands) is used in abscissa instead of the effective temperature (or the spectral type). The HRD is then referred to as the color-luminosity or colour-magnitude diagram if the luminosity is replaced by the absolute magnitude in a specific band. The advantage is that one can incorporate stars that are observed only photometrically and not spectroscopically. \\

The majority of stars spend their life on the so-called {\it Main Sequence} (black box in the right panel of Fig.~\ref{chap1:fig4}). It is represented in the diagram by a diagonal band running from the top left (hot massive stars) to the bottom right (cool and low-mass stars). The Main Sequence is the phase in which stars stay while burning hydrogen in their core, i.e., the majority of their lives. About 90\% of all the stars lie on the Main Sequence, including the Sun. The Sun is located at $T_{\rm eff} = 5780$~K and with a luminosity in the logarithmic scale of $\log \left(L/L_{\odot} \right) = 0$ (corresponding to a visual magnitude $M_V = 4.81$ mag, \citealt{willmer18}). \\

In the lower left part of the HRD are the {\it white dwarfs}. These are remnants of stars that were once similar to the Sun but have shed their outer layers and no longer undergo fusion. They are very hot but have low luminosities due to their compact sizes. 

Above the Main Sequence, in the upper right part of the HRD are the {\it giants} and {\it supergiants}. These stars have evolved off the Main Sequence and have expanded to large sizes, resulting in high luminosity but cooler surface temperatures. Supergiants (and to a lesser extent giants) have exhausted all their hydrogen into their core and are undergoing helium burning into their core. Their helium core is still surrounded by a hydrogen shell. \\

On the top left of the Main Sequence, among the hot, massive stars, we find the classical {\it Wolf-Rayet} (WR) stars \citep{crowther07}. These stars are evolved massive stars that have peeled off their hydrogen-burning shell due to their powerful winds, and have an exposed helium core. Although O and WR stars feature heavy mass-loss rates, the latter, when integrated over the stellar lifetime are usually not sufficient to enable a direct evolution between these two evolutionary phases. At some point in their past, the progenitors of WR-stars must therefore have undergone a phase of extreme mass loss, during which their outer envelopes were removed to reveal the bare core that became the WR star. This extreme mass loss is thought to occur either during a {\it red supergiant} (RSG) or {\it Luminous Blue Variable} (LBV) phase \citep{maeder00}. Therefore, according to the classical evolution scenario (called as Conti scenario, \citealt{conti76}), once a massive star with an initial mass higher than 25~\msun\ leaves the Main Sequence, it will first evolve into a phase of red supergiant (RSG) or luminous blue variable (LBV), before reaching the stage of WR stars. As soon as a star becomes a WR star, it will first evolve into a nitrogen-rich stage (WN), before going into a carbon-rich stage (WC), and ending into a oxygen-rich stage (WO), before exploding or collapsing as supernova. In order for an O star to evolve into a WR star, the O star must thus lose a sizable fraction of its mass. 

The HRD can thus be used as a nice tool to probe the evolution of the stars. It is particularly important for understanding stellar populations, their distributions, their formation history, and for estimating their ages. However, one must keep in mind that theoretical predictions are computed based on initial parameters such as the metallicity, the convective core overshooting, and mass-loss prescriptions\footnote{Convective core overshooting refers to the phenomenon where material in a star’s convective core moves beyond the boundary predicted by traditional models of convection. This provokes the mixing of additional material into the core of the star, impacting the lifetimes and the luminosities of the stars, nucleosynthesis, and the formation of remnants like white dwarfs, neutron stars or stellar-mass black holes. This occurs because of the inertia of rising and sinking convective material, allowing it to penetrate into the stable radiative zone surrounding the convective core. The overshooting extends the effective size of the convective core, extends the stellar lifetime, and altered the chemical profiles\\ Mass loss prescription refers to the mass and angular momentum that is lost by a star through its winds. The amount of mass lost depends on the type of the star, its initial mass, its age, its metallicity and its evolutionary stage. This will affect the structure of the star and its evolutionary path. The ejected material will enrich and reshape the medium surrounding the star, providing a new material for the next generation of stars and planets.}. Therefore, these predictions cannot be trusted blindly, as they require observational validations from basic stellar observables.

\section{Fundamental stellar parameters}\label{chap1:sec2}

\subsection{Mass}\label{sec2:subsec1}
%Basic observables allow astronomers to estimate a star's mass and age. For instance, the mass of a star determines its lifespan and the path it will take through various stages of stellar evolution, from birth in a nebula to potentially ending as a white dwarf, neutron star, or black hole.
Mass is the prime parameter that governs the fate of a star. Any modification of this parameter over the life of a star, through stellar winds, mass exchange with a companion, or collision, will have tremendous consequences over the way the star evolves. 
Although the stellar mass is of paramount importance to understand how a star evolves, this parameter is extremely difficult to constrain, and cannot be directly measured. When a star is considered as single, there are two different ways to derive its stellar mass: 1) by comparing its position in the HRD (see Sect.~\ref{sec2:subsec6}) with theoretical evolutionary tracks (a.k.a. evolutionary mass), 2) by inferring the stellar mass through an estimation of the surface gravity of the star and its radius (a.k.a. spectroscopic mass). However, these two methods rely on stellar or evolution models, and have therefore caveats. Indeed, the evolutionary tracks are computed from initial input parameters such as convective core overshooting, wind prescriptions, rotation, and metallicity \citep{martins13}. These input parameters can have a severe impact on the predicted masses, and ages and their accuracy, which can lead to different estimations for the stellar mass. It is a long-standing problem among massive stars, that is called {\it mass discrepancy} \citep{herrero92, martins12, mahy15, tkachenko20, mahy20}.

Spectroscopic analyses are performed by fitting observed spectra with synthetic spectra computed with stellar atmosphere codes (e.g., MARCS, \citealt{gustafsson75,gustafsson08}; PHOENIX, \citealt{hauschildt99}; ATLAS, \citealt{castelli03}; TLUSTY, \citealt{hubeny88}; CMFGEN, \citealt{hillier98}; Fastwind, \citealt{puls05}; or PoWR, \citealt{hamann03}, among others). To obtain the spectroscopic mass $M_{\rm spec} = g~R^2/G$ (with $G$ the gravitational constant), the surface gravity ($\log g$), and the radius of the star must be derived accurately. It is therefore crucial to derive the luminosity and the effective temperature of the star accurately (see Sect.~\ref{sec2:subsec2}). For high-mass stars, the surface gravity is derived from the broad Balmer lines. These lines are hard to normalize, which can introduce a systematic bias in the determination of the surface gravity. The line broadening due to the projected rotational velocity (v sin i, see Sect.~\ref{sec3:subsec6}) and other velocity fields at the surface (known as macro-turbulence) may also affect the determination of the surface gravity. For stars with strong winds like evolved massive stars (WR stars, or LBVs), the spectral lines may appear completely in emission or the core of the absorption lines can be filled by emission features, preventing any determination of the surface gravity. Moreover, their distance and therefore their luminosity can also be poorly determined. All in all, the reliability of the masses derived through these methods can lead to $15-20$\% of uncertainties. \\

For low- and intermediate-mass stars, other techniques have been used to estimate their masses but these techniques also reach uncertainties of the order of 15\% of the masses. These methods are (1) H$\alpha$ fitting \citep{Bergemann16}, (2) C/N ratio \citep{ness16,martig16}, and (3) Li abundances \citep{DoNascimento09}. They have been summarized in the review of \citet{serenelli21}. We refer the reader to those papers for a complete description of these techniques.\\

The only way to achieve a better accuracy on the stellar masses is by directly measuring them from binary systems. About 50\% of low-mass stars and more than 70\% of high-mass stars are members of binary or multiple systems. By applying Kepler's third law $$\frac{G(M_1+M_2)}{4~\pi^2} = \frac{a^3}{P_{orb}^2}$$ where $G$ is the gravitational constant, $M_1$ and $M_2$ the masses of the primary and secondary, respectively, $a$ the semi-major axis and $P_{orb}$ the orbital period of the system, we have a unique chance to measure the masses of stars, independently of models and calibrations. Spectroscopy gives the possibility to measure radial velocities of the components but prevents from deriving the inclination of the system. Therefore, only minimum masses can be estimated for each component. By combining spectroscopy with photometry or astrometry, the inclination of the system can be determined, yielding to the absolute individual masses of the components, with potential accuracies of 1\% \citep{torres10}.

\subsection{Luminosity}\label{sec2:subsec2}

The luminosity is defined by the total energy emitted by a star per second. If we know the effective temperature and the radius of a star, its luminosity can be directly measured through the relation:
\begin{equation}\label{eq1}
L = 4\pi~\sigma~T_{\rm eff}^4~R^2, 
\end{equation}
where $\sigma =5.669 \times 10^{-5}$~erg~s$^{-1}$~cm$^{-2}$~K$^{-4}$ is the Stefan-Boltzmann constant.

However, for isolated stars, directly measuring the stellar radius of an individual object is not an easy task (see Section~\ref{sec2:subsec4}). Although the luminosity can be derived from the bolometric magnitude, this requires to know with precision the distance of the star, its extinction, and its bolometric correction. The uncertainty on the luminosity is often dominated by the uncertainty on the distance. For those objects, the luminosity is derived following these relation: 
\begin{equation}
    M_X = X - A_X - (5 \log_{10} \left(d \right) - 5)
\end{equation}
where $X$ is the apparent magnitude in a specific band, $A_X$ is the extinction in the same band, and $d$ is the distance of the star. The bolometric magnitude $M_{\rm bol}$ is deduced by the formula:
\begin{equation}
    M_{\rm bol} = M_X + {\rm BC}_X
\end{equation}
where ${\rm BC}_X$ is the bolometric correction in the $X$ band.\\
From $M_{\rm bol}$, we compute the luminosity using:
 \begin{equation}
    \log \left(L/L_{\odot} \right) = (M_{\rm bol} - 4.74)/-2.5
\end{equation}
where $4.74$~mag is the bolometric magnitude of the Sun \citep{mamajek15}.\\

The luminosity is also dependent on the mass of the star. There exists a relation between stellar mass and stellar luminosity. This relation has been determined thanks to the observations of members of binary systems. For stars with initial masses higher than 0.5~\msun, the mass and the luminosity follow the relation:
\begin{equation}
    L/L_{\odot} \sim (M/M_{\odot})^\alpha, 
\end{equation}
where $L_{\odot} = 3.826 \times 10^{26}$~erg~s$^{-1}$ and $M_{\odot} = 1.989 \times 10^{30}$~kg and represent the luminosity and the mass of the Sun, respectively. It has been shown that this relation is not the same through all the mass ranges, and the exponent $\alpha$ can vary from about 2 to 6, even though in the massive star regime, very few stars have been thoroughly analyzed so far \citep{torres10,eker18,mahy20}. 

\subsection{Effective temperature}\label{sec2:subsec3}

Since we see near black-body emission from the photosphere when we view a star, we commonly use the photospheric temperature of a star to characterize it. The effective temperature $T_{\rm eff}$ of a star is defined as the temperature of the black-body emitting exactly the same amount of radiative energy as the star. This means that the effective temperature is similar to the temperature of the star at the point where the visible flux is emitted, which is in general at or close to the photosphere. The effective temperature depends on the color of the star and its spectral type, with hotter stars appearing bluer and cooler stars redder (see Sect.~\ref{sec1:subsec1}). The Sun has an effective temperature of $\sim$~5780~K, which means that it will appear yellow to the human eye (see Fig.~\ref{chap1:fig1}).\\

The temperature in the atmosphere of a star is not constant, it decreases from the interior (where the energy is produced) towards the outer region of the atmosphere (where the energy is released). The exact profile of the temperature structure depends on the detailed physics of the atmosphere, especially the opacities. The opacity is a measure of how transparent or opaque the material of a star is to the passage of the radiation. That depends on its composition, density, and temperature. In cool stars, the opacity is often dominated by molecules such as \ce{H2O} or \ce{TiO} or neutral hydrogen atoms and \ce{H-} ions, which are very effective at absorbing radiation. In hot stars, the opacity is dominated by free electrons and the absorption of light by ionized helium and other heavier elements. The higher temperatures ionize most elements, leading to processes like electron scattering and {\it bound-free} transitions.\\

The role of the stellar atmosphere is to transport the energy outward. In massive stars, this is done by radiative transfer, which depends on the temperature gradient: the higher the temperature gradient, the more efficient the transfer of energy. In cool stars, the opacity is high, photons are absorbed and re-emitted frequently, such as the radiative transport is inefficient. In such cases, the energy transport shifts to convection, where hot material moves outward and cooler material sinks inward. For main-sequence stars, as the effective temperature increases, the opacity decreases, which can influence the structure of the stars. In order to maintain an efficient energy transport, a balance between temperature gradient and opacities must be found. Stars with $M < 0.5~\msun$, classified as M-type stars, or later (L-, T-, etc), are relatively cool, leading to higher opacity in their outer layers. They have deep convective envelopes and sometimes even fully convective interiors. Sun-like stars ($0.5~\msun < M < 1.5~\msun$, or spectral types K, G or F) have a radiative core where energy is transported primarily by radiation and an outer convective zone where opacity increases due to the ionization of hydrogen and helium, leading to convection. Finally, hotter stars (those with spectral types O, B, or A) have a radiative envelope and a convective core (see Sect.~\ref{sec1:subsubsec3}). % This explains why the inclusion of metals may significantly alter the temperature structure of massive stars at fixed effective temperature. It is thus crucial to understand to which extend the inclusion of metals in massive stars atmospheres modifies the temperature distribution, since such a change will imply modifications of the ionisation structure, thus of the spectral lines and consequently of the spectral types. In other words, the relation between spectral-type and effective tempera- ture may be changed.

\subsection{Radius}\label{sec2:subsec4}
Although fundamental, the radius of a star is not a well-defined quantity and requires sophisticated techniques to measure it. The first possibility is by knowing the luminosity and surface temperature of the star through Equation~\ref{eq1}. However, as already mentioned, these quantities require a good knowledge of the distance of the stars. A second way to measure the stellar radius consists by using optical interferometry to measure the angular diameter of a star. This technique only works for nearby stars, and combines the light from multiple telescopes to create a single, high-resolution image that can be used to measure the angular size of the star. Once the angular diameter ($\theta$) is measured and the distance of the star ($d$) is known, the physical radius ($R$) can be calculated by the formula:
\begin{equation}
    R = \frac{\theta~d}{2}
\end{equation}
This technique has notably been used to measure the radius of the red supergiant Betelgeuse \citep{michelson20}.\\

A third technique aims to apply the method of transit. This can be done for planets, but also for stars, when they are in eclipsing binary systems. In these systems (see Sect.~\ref{sec4:subsec2} for more details), two stars orbit each other such that one periodically passes in front of the other in the sight-line of the observer. The total brightness of the system will drop during the eclipses which allows the astronomer to directly measure the radii of both stars. 

The accuracy of these different methods is not the same. While a direct measurement through interferometry or photometry can give accuracies of the order of about a percent, the first technique will give radii with an accuracy of a few percent depending or not whether we know the distance of the star with high precision.

\subsection{Distance}\label{sec2:subsec5}
For centuries the problem of stellar distances has puzzled astronomers, although the underlying geometric principle needed to ascertain them is extremely simple and well understood. Knowing precisely the distances of the stars is crucial to accurately derive their stellar parameters. It also essential for mapping the three-dimensional structure of our Galaxy, understanding the dynamics of the stars, and to larger scales, for providing evidence for the expansion of the universe. 

In December 2013, the satellite Gaia of the European space agency ESA was launched with the aims of revolutionizing our knowledge of the Milky Way. Gaia was designed to provide a quantitative census of the stellar populations, to clarify the origin and history of our Galaxy and to perform extremely accurate astrometric measurements to derive proper motions and parallaxes of a significant part of stars in the Galaxy. Since the advent of the Gaia mission, the measurement of the stellar distances has improved significantly over the last 10 years. The first and second data releases (DR) of the mission, DR1 and DR2, took place on September 14th, 2016 and on April 25th, 2018, respectively.\\

This satellite measures the parallaxes of the stars with unprecedented precision, i.e., the angle subtended at the star by one astronomical unit (i.e., 150 million km) or half the apparent diameter of the Earth orbit when seen from the star. It uses the motion of the Earth around the Sun to measure the distance of the stars by triangulation. You can experiment by moving your finger to different distances in front of your face and blinking from eye to eye. You will see that your finger appears to move a greater distance when it is closer to your face than when it is further away. Gaia processes the same way. When an observer moves, the nearby objects appear to shift position relative to more distant objects. As Earth orbits the Sun, astronomers measure the angle through which the stars appear to move over the course of the year, and knowing how far the Earth is moving (Earth's large distance to the Sun, i.e., one astronomical unit), astronomers are able to calculate the star's distances. The third Gaia data release is more recent (June 13th, 2022) and allows us to determine accurate distances and proper motions for more than 1.8 billion stars. To this date, the Gaia mission represents the biggest census of objects in our Milky Way.\\

Certain types of stars, like Cepheid variables, have predictable luminosities. By observing their apparent brightnesses, astronomers can determine their distances. This class of stars gathers very bright radially pulsating stars (i.e., stars whose layers periodically expand and contract with the whole layer moving in phase) that are located in a specific region of the HRD. Classical Cepheids are Population I F-K supergiants with effective temperatures between 5000 and 7000 K, and luminosities of order $10^4~L_{\odot}$. These objects are evolved stars that are burning helium in their core. Their progenitors were B-type stars with masses between 3 and 16~\msun. The majority of Cepheids are observed during their second crossing of the instability strip which lasts longer than the first crossing. In 1912, the american astronomer Henrietta Leavitt (1868 - 1921) studied a sample of Cepheids in the Large Magellanic Cloud. As these stars were in the same distant cloud they were all at much the same relative distance from us. Therefore any difference in apparent magnitude could be translated into a difference in absolute magnitude (see Sect.~\ref{sec1:subsec2} for more details). She thus noticed that Cepheids with longer pulsating periods were brighter than those with shorter periods, establishing for the first time the existence of a relation between their pulsation period and their intrinsic luminosity. Five years later, Harlow Shapley, an American astronomer, used a larger sample of Cepheid stars to re-calibrate the absolute magnitude scale for Cepheids and to revise the value of the distance to the Small Magellanic Cloud. He also studied Cepheids to estimate the distances of a large sample of globular clusters in the Milky Way he observed. Although his studies focused on stars that were wrongly believed to be classical Cepheids, he established a first measurement of the size of the Milky Way, and demonstrated that the Sun was not located at the center of our Galaxy, but off by 50,000 light years. These results were then used by Edwin Hubble (1889 - 1953) to prove that M31 was a galaxy, and not a nebula inside the Milky Way. Since then, the calibration of the period-luminosity relation has been revised many times and is generally parameterized with $\alpha$ and $\beta$ as follows:
\begin{equation}
   \langle M_X \rangle = \alpha (\log~P - p_1) + \beta + \gamma([\ce{Fe/H}] - p_2),
\end{equation} 
 where $M_X$ is the absolute magnitude in the $X$ band, $P$ is the pulsation period, $p_1$ and $p_2$ are constants \citep[see ][and references therein]{groenewegen24}, and [\ce{Fe/H}] is the logarithmic ratio of the iron-to-hydrogen ratio, a commonly used abbreviation for metallicity, that is expressed as $\left[Fe/H\right] = \log_{10} \left((N_{Fe}-N_{H})/(N_{Fe}-N_{H})_{\odot} \right)$, with $N_{Fe}$ the number of iron atoms, and $N_{H}$ the number of hydrogen atoms in the object. 
 
 Many studies tried to revise this relation using Gaia. However, one of the main issues with Gaia was to correct for the zero point parallax because point sources located at almost infinity do not have a Gaia parallax of zero. Several papers tried to correct for that with more or less success \citep[see e.g., ][]{lindegren21, groenewegen21,maizapellaniz22}. Another question that has been raised regarding the period-luminosity relation is its possible metallicity dependence. Using a period-luminosity relation calibrated with metal-rich Cepheids to measure the distance to a galaxy with predominantly metal-poor Cepheids could lead to an overestimation of the distance.

For more distant galaxies, astronomers rely on the exploding stars known as supernovae. Like Cepheids, the rate at which a certain class of supernovae brightens and fades reveals their true brightness, which then can be used to calculate their distance. But this technique also requires good calibration using the distance determination of Cepheids. Without knowing the precise distances to a few (rather close) supernovae, there is no way to determine their absolute brightness, so the technique would not work.
%
%Measuring the distance to stars is crucial for mapping the three-dimensional structure of our galaxy. Techniques like parallax help establish the distances to nearby stars, providing a foundation for understanding the scale and structure of the Milky Way.

%Motion and Velocity: Observing the motion of stars (proper motion and radial velocity) helps in understanding the dynamics of stars within galaxies, including the rotation of the Milky Way and interactions between stars.

%Standard Candles: Certain types of stars, like Cepheid variables, have predictable luminosities. By observing their apparent brightness, astronomers can determine their distance. This is crucial for measuring distances on a cosmic scale and understanding the expansion of the universe.

%Redshift and Hubble's Law: The motion of stars and galaxies, observed through redshift, provides evidence for the expanding universe and is foundational for cosmology.

\section{Inferred properties}\label{chap1:sec3}

\subsection{Chemical composition and metallicity}\label{sec3:subsec6}

\begin{figure}[b]
\centering
\includegraphics[width=12cm,trim=40 300 800 30,clip]{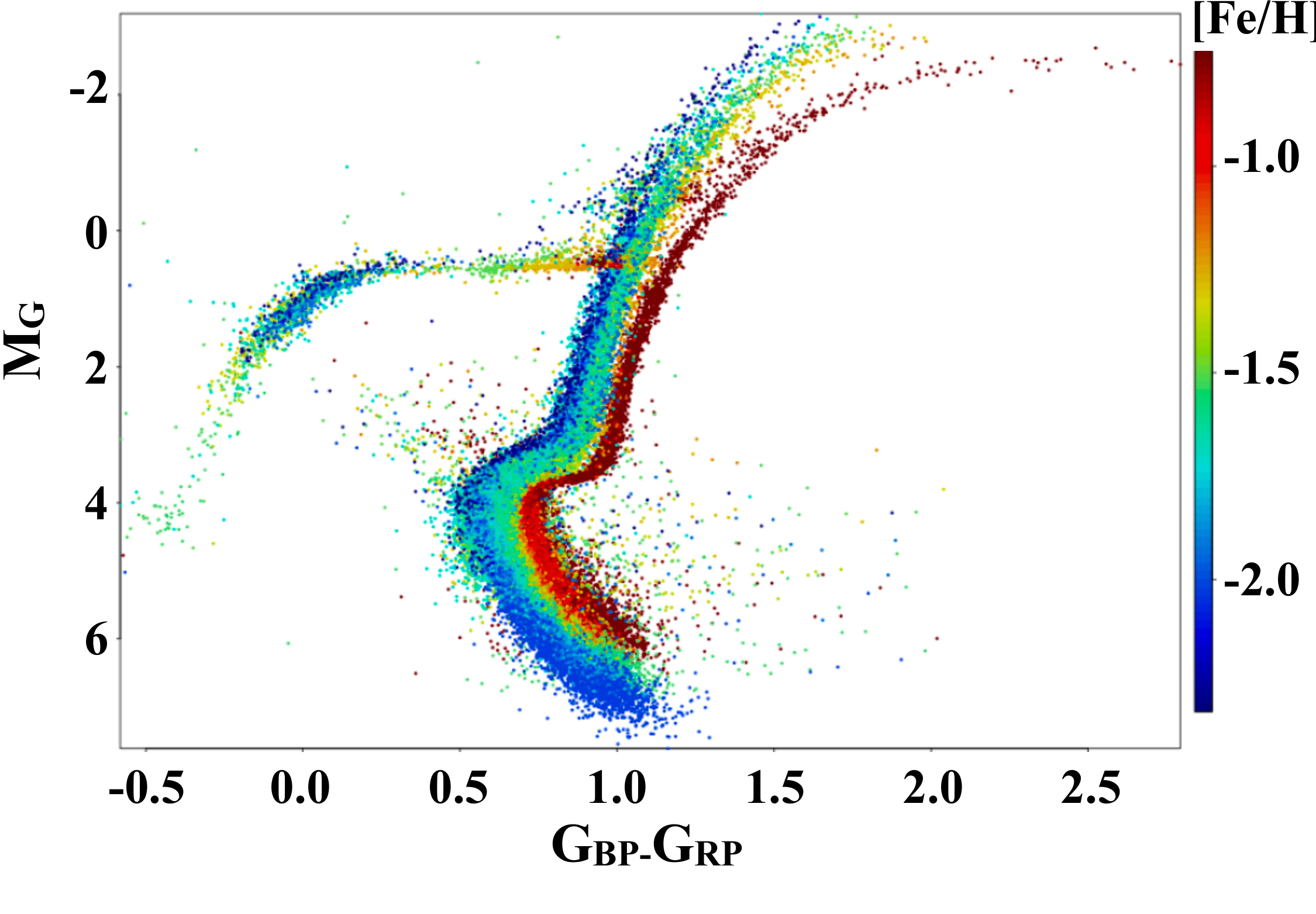}
\caption{Observational Color-Magnitude diagram of 14 globular clusters constructed on the basis of measurements obtained through the second Gaia data release \citep{babusiaux18}. The color-coding is according to metallicity.}
\label{chap1:fig5}
\end{figure}

The chemical composition of stellar matter is of paramount importance to describe the evolutionary scheme of the stars. It indeed determines the basic characteristics such as radiation and energy production due to the nuclear reactions. These reactions, in turn, change the chemical composition of the star. 

The stellar composition of stars is often defined by the parameters $X$, $Y$, and $Z$ such as $X+Y+Z=1$, where $X$ represents the mass fraction of hydrogen, $Y$ the mass fraction of helium, and $Z$ the mass fraction of all the remaining elements, called as "metals". For a main-sequence star in our Milky Way, $X$ is found to be between 0.68 and 0.73, and $Y$ between 0.24 and 0.29. The mass fraction of heavy elements, on the other hand, varies strongly from star to star, and lies in between $Z = 10^{-6}$ to about $Z = 0.04$. When the Universe was created, stars were only made of hydrogen and helium. Many heavier elements are created by nucleosynthesis in stars. Therefore, when the stars are on the Main Sequence or during their late evolutionary stages, a large fraction of their mass is ejected into the interstellar medium, either through powerful stellar winds, or during supernova explosions. This has a considerable influence on the environments of the stars both mechanically and through radiation as it will enrich their surroundings with heavy chemical elements, and reshape the interstellar medium. This newly-enriched material will then be incorporated into the next generation of stars (and planets) that will be formed in this medium, making them more rich in metals (Fig.~\ref{chap1:fig5}). Consequently, the broad range of $Z$ values must be interpreted as a broad range of stellar ages. The stars with very low $Z$ are the first-generation stars which were formed before significant chemical enrichment of the interstellar medium took place, i.e., just after the Big Bang. 

Stars in our closest satellite dwarf galaxies, the Large and Small Magellanic Clouds (LMC and SMC, respectively), are of particular interest to understand the role of the metallicity on the stellar evolution due to their lower metal content. The LMC is indeed expected to have a metallicity close to half of the solar metallicity ($Z_{\odot}$), while the SMC metallicity is close to 1/5$^{\rm th}$ $Z_{\odot}$. These environments offer the astronomers the opportunity to study stars in conditions very close to the creation of the first stars and the early universe. 

%\subsection{Stellar Age}\label{sec3:subsec7}

%\subsection{Wind parameters}\label{sec3:subsec8}

\subsection{Rotation}\label{sec3:subsec9}

\begin{figure}[b]
\centering
\includegraphics[width=12cm,trim=30 340 50 45,clip]{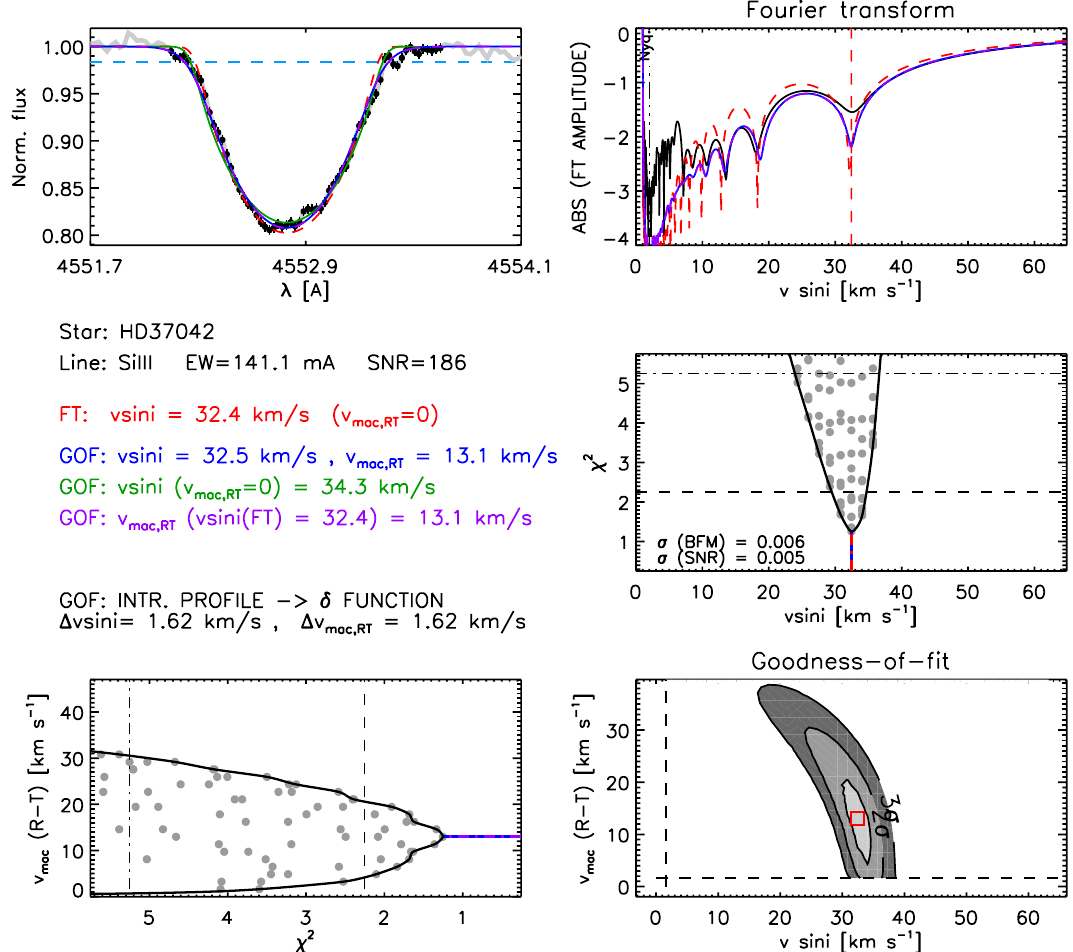}
\caption{Example of output from the {\textsc iacob-broad} tool \citep{simondiaz14} that aims at measuring the projected rotational velocity of stars. Five main graphical results are presented: (a) the line profile (upper left); (b) the Fourier Transform of the line profile (upper right); and (c) 2D $\chi^2$-distributions resulting from the Goodness-of-Fit analysis (lower right) and their projections (middle and lower left). We refer the reader to the dedicated paper.}
\label{chap1:fig7}
\end{figure}

Observed spectral lines are always broadened. This is due to the finite resolution of the spectrographs and is also intrinsic to the star. Studying the width of spectral lines provides information about the various physical processes in the atmosphere of the star:
\begin{itemize}
    \item Thermal broadening: caused by the motion of the ions due to thermal energy,  
    \item Pressure broadening: caused by the collisions between particles in the stellar atmosphere that affect the energy levels of these ions.
    \item Rotation broadening: caused by the rotation of the star, 
    \item Turbulence due to the turbulent motion in the stellar atmosphere.
\end{itemize}

It is possible to measure the projected rotational velocity ($v~\sin~i$) of the star from the spectral lines. The lines of a rotating star are broadened because light from the limb that approaches Earth is blue-shifted while the light of the receding limb is red-shifted. To measure the projected rotational velocity, astronomers apply Fourier transform on dedicated spectral lines, with a preference for metallic lines \citep{carroll33, gray76, gray05, simon-diaz07}. This technique is based on the identification of the first zero crossing in the Fourier transform of the line profile (Fig.~\ref{chap1:fig7}). The corresponding spatial frequency $\sigma_1$ is then converted into $v~\sin~i$ through the relation:
\begin{equation}
    \frac{\lambda}{c}~v~\sin~i~\sigma_1 = 0.660,
\end{equation}
where $\lambda$ is the central wavelength of the spectral line, and $c$ is the light speed.\\

When a star is single, it will conserve its angular momentum as it evolves. For instance, a star that contracts during its life will increase its rotation rate, while a star that expands will decrease its rotational rate. More evolved stars thus have lower rotational velocities than stars close to the Zero-Age Main Sequence. Stellar rotation is moreover important because it affects the internal structure of the stars, and therefore their evolution. Indeed, faster rotation increases the mixing efficiency in the core layers of stars. It brings fresh hydrogen into the core and transports heavier elements to the surface, which can prolong the lifespan by delaying hydrogen depletion. The stars can therefore appear younger than they are in reality. Faster rotating stars will be more enriched in heavier elements at their surface (nitrogen being the most important by-product of the CNO cycle), that is what we call rotational mixing \citep{maeder00}. Rotational mixing arises because stars are not rigid bodies; different layers can rotate at different speeds (differential rotation). This creates complex flows within the star that mix material from different regions. The two main physical mechanisms behind this process are the shear instability and the meridional circulation. The shear instability is due to differential rotation. In fast rotating stars, the equator rotates faster than the poles or deeper layers rotate at different speeds than the outer layers developing velocity gradients under the surface. These gradients can cause turbulent mixing between layers of different velocities. This allows elements like hydrogen, helium, and heavier elements to move between layers. The meridional circulation is described by large-scale circulatory flows within the star, similar to global atmospheric circulation on Earth. Material from the equator moves toward the poles along the surface, and material from the core moves toward the surface or equator in deeper layers. This circulation drives the mixing of elements between the star's core and outer layers. Rotational mixing is more prominent in massive stars because they tend to rotate more rapidly. In lower-mass stars, this process is weaker, but still can play a role. \citet{Hunter08} studied the nitrogen enrichment of a large sample of apparently single stars in the LMC. When plotting the surface nitrogen enrichment against the projected rotational velocity of the stars, they observed a general trend between these two quantities, proving the existence of rotational mixing. However, two groups were not explained by theory: (1) a group of slowly rotating stars with nitrogen excess at their surface, (2) a group of fast rotators with no surface enrichment in nitrogen. Although these two groups are still unexplained, several hypotheses have been advanced to explain their existence. The first one was that these stars have been members of binary systems, that suffered from binary interaction and that are now disrupted. Many massive stars exist in close binaries, that can undergo mass and angular momentum transfers. During these interactions, mass and angular momentum could be lost by one star and transferred to the companion. This process slows down the donor star, and all the material that has been removed from its surface exposes the deeper layers of the star, more enriched in heavier elements such as nitrogen. The material removed from the donor is accreted by the companion, increasing its mass and its angular momentum. The mass gainer looks thus like a fast rotator with no clear surface enrichment. Magnetic fields, binary mergers, inefficient mixing processes have also been proposed among others as alternative explanations to form these two abnormal groups, but no clear consensus has been reached yet.

\section{Binarity}\label{chap1:sec4}

Most of the stars belong to binary or higher-order systems \citep{moe17}. When the companion orbits at a sufficiently wide separation from its host star, the evolution of the stars is not influenced by their companion. Such {\it wide} binaries follow the evolutionary theory of single stars. However, about 50\% of low-mass stars, and more than 70\% of high-mass stars are found to have a close-by companion with whom the host star will interact during its lifetime. In the massive star regime, these interactions dominate the evolution through mass and angular momentum transfer, envelope stripping or coalescence \citep{sana12}.\\

\begin{figure}[htbp]
\centering
\includegraphics[width=12cm,trim=250 0 0 10,clip]{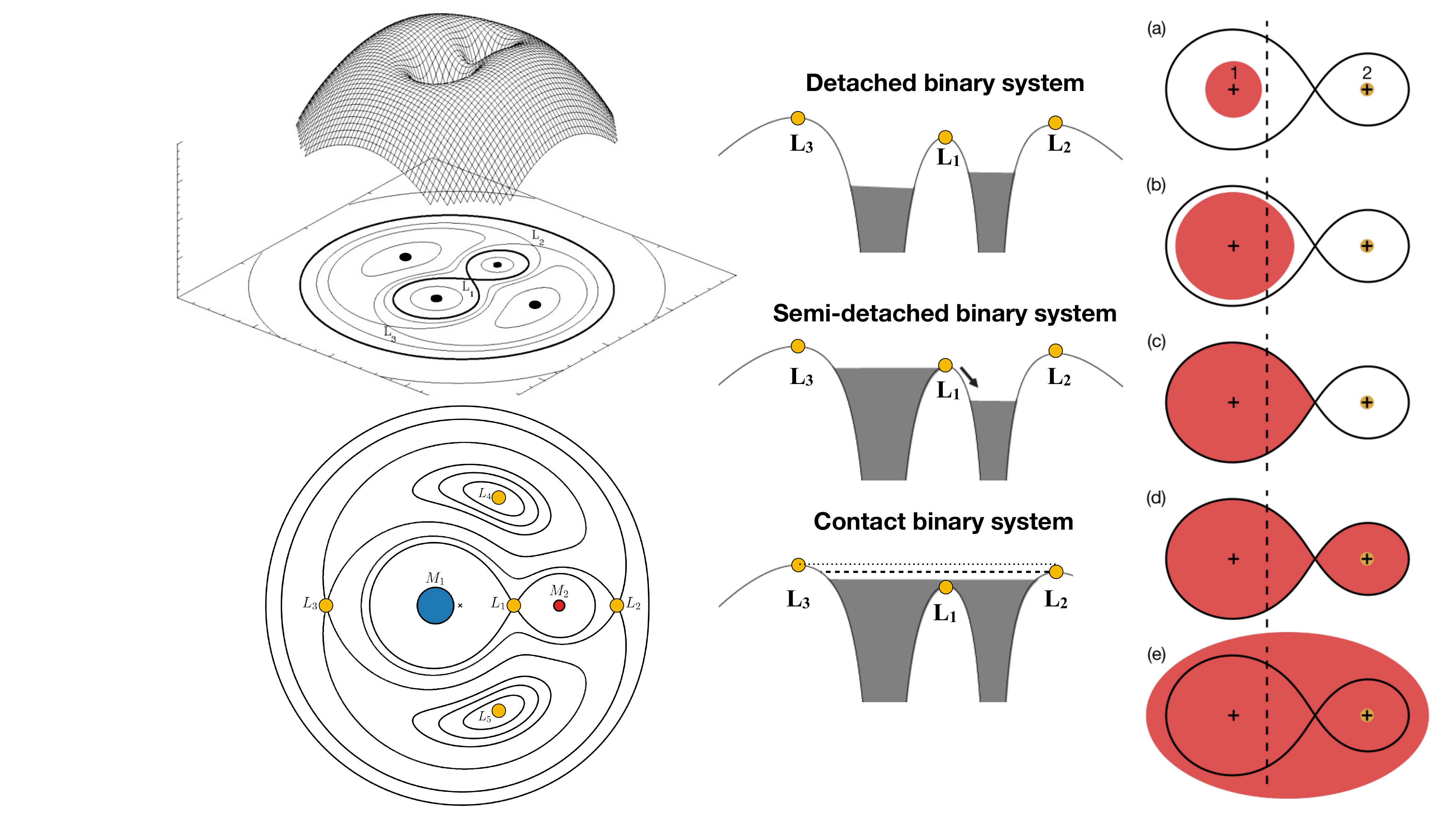}
\caption{Representation of the Roche Lobe representation in 3D (upper left) and in 2D (lower left). A vertical cut along the Roche lobes is shown in the middle panel during the three different configurations of the system. The right panel represents the evolutionary scheme of a binary system - (a) detached system, (b) primary star almost fills its Roche lobe, (c) primary star fills its Roche lobe, transferring its material to its companion, (d) secondary star also fills its Roche lobe, (e) over-contact phase. }
\label{chap1:fig6}
\end{figure}

Binary systems are genuine astrophysical laboratories, and constitute the only direct way of measuring the mass (and to a lesser extent the radius) of a star. Let us consider two stars that are gravitationally bound and orbit each other closely on a circular orbit. One defines the Roche lobe as the region around each star within which the material is gravitationally bound to that star. The shape of a star in a binary system will be ruled by the Roche potential. This potential $\Phi$ is the sum of three components: (1) the gravitational potential of the primary star, (2) the gravitational potential of the secondary star, and (3) the potential due to the centrifugal force that arises because the system is moving around its center of mass. It can be written:
\begin{equation}
    \Phi (x,y,z) = - \frac{GM_1}{r_1} - \frac{GM_2}{r_2} - \frac{1}{2} \omega^2 (x^2+y^2),
\end{equation}
where $G = 6.674 \times 10^{-11}$~m$^3$~s~kg$^{-1}$ is the gravitational constant, $M_1$ and $M_2$ are the masses of the primary and secondary, respectively, $r_1$ and $r_2$ are the distances from the point $(x,y,z)$ to each star, $\omega$ is the angular velocity of the system's rotation, $x$ and $y$ are the coordinates in the plane of rotation, with the origin typically at the center of mass of the system. If we choose the orbital separation between the stars $a$ as the unit for distance and adopt the center of star 1 as the origin of our frame of reference, the potential can be written in a dimensionless form: $\Omega = (-\Phi~a)/(G~M_1)$. \\

The Roche potential features five points where the gravitational and centrifugal forces exactly balance, the so-called {\it Lagrangian points}. Their position can be seen in Fig.~\ref{chap1:fig6}, and obtained by solving the equation $\bigtriangledown \Omega = 0$. The first three Lagrangian points $L_1$, $L_2$, and $L_3$ are also called the libration points. \citet{Eggleton83} give an approximate analytical expression for the radius of a sphere that would have the same volume as the Roche Lobe:
\begin{equation}
    \frac{R_{\rm RL}}{a} = \frac{0.49~q^{2/3}}{0.6~q^{2/3} + \ln(1+q^{1/3})}
\end{equation}
where $q$ is the mass ratio, and $a$ the separation between the two stars. \\

The Roche lobe represents a physical limit to the expansion of a star in a binary system. While a single star can expand to very large radii, the volume of a star in a binary system is limited to $\frac{4 \pi}{3} R^3_{\rm RL}$. The Roche potential is critical for understanding how stars interact with their companions and how mass and angular momentum are transferred between them in binary systems. When a star evolves, it expands to fill its Roche lobe. The material can then flow through the $L_1$ point towards the companion star, leading to the mass transfer (see the middle panel of Fig.~\ref{chap1:fig6}). This conducts to different configurations of the system. When none of the stars is filling their Roche lobe, the system is called {\it detached}. When the most massive star fills its Roche lobe, transferring its material to its companion, the system is called {\it semi-detached}. Finally, when both stars fill in their Roche lobe, the system is called in {\it (over-)contact}\footnote{A contact binary system consists of two stars that have filled their Roche lobes and started sharing a single envelope, a over-contact binary system are two stars forming a binary system that share a single envelope but overfill their Roche lobes.} (see the right panel of Fig.~\ref{chap1:fig6}). \\

To observationally detect binary systems, techniques such as spectroscopy, photometry, interferometry or astrometry can be used but they will not probe the same configurations nor the same separation range between the components. Spectroscopy and photometry will probe shorter-period orbits, while interferometry and astrometry will allow us to detect longer period orbits.  

\subsection{Spectroscopic binaries}\label{sec4:subsec1}
Spectroscopic binaries are systems detected through the periodic variation of the wavelengths of the spectral lines. When the orbital plane of a binary system is not perpendicular to the line of sight, the components of the system will move alternatively towards and away from the observer. When one component moves towards the observer, its spectral lines are blue-shifted (i.e., shifted to shorter wavelengths), whereas simultaneously the lines of the other component move away from the observer, and are red-shifted (i.e., shifted to longer wavelengths). By measuring the shift between the observed wavelength ($\lambda$) and the rest wavelength of a spectral line ($\lambda_0$), it allows us to determine the radial velocity along the line of sight towards the observer:
\begin{equation}
    \frac{\lambda - \lambda_0}{\lambda_0} = \frac{v_{\rm rad}}{c}, 
\end{equation}
where $v_{\rm rad}$ is the radial velocity of the star and $c$ is the speed of light. \\ \\
Spectroscopic binaries can be classified in different categories depending on the number of components one can see in the spectra:
\begin{itemize}
    \item Double-lined spectroscopic binaries (SB2) where both components have a comparable brightness and the lines of both stars are seen in the spectrum, moving anti-phase with respect to each other.   
    \item Single-lined spectroscopic binaries (SB1) where only one component can be identified in the spectrum. Most of the time the star that can be measured is more luminous (and often also more massive) than its companion. The companion may also be a compact object, that does not contribute to the total light of the system such as a neutron star, or a stellar-mass black hole. Some objects classified as SB1 in the literature are false detections. In general, they are misclassified as SB1 because of their small RV variations originated from radial or non-radial pulsations or other motions in the atmosphere of the star,
    \item it is also possible to classify multiple-order systems as SB$_i$ when $i$ components are observed in the spectrum. 
\end{itemize}

The analysis of the spectral line shifts provides the radial velocities of the components which allows the measurements of orbital parameters such as the orbital period of the system, its eccentricity, the longitude of the periastron, the velocity of the center of mass of the system and the projected size of its semi-major axis. The only parameter that is not possible to determined through spectroscopy is the inclination of the system. Therefore only minimum masses can be derived for the components through this technique.

\subsection{Eclipsing binaries}\label{sec4:subsec2}
Photometry allows us to study the variation of the total light of a star as a function of time. When two stars are orbiting each other and the orbital plane of the system is aligned such that one star periodically passes in front of the other, the system is called eclipsing. The shape, the depth and the timing of the eclipses in the light curve provide detailed information about the configuration of the system (detached, semi-detached or contact), the orbital period, the eccentricity, the inclination, and also the relative radii of the stars. By combining photometry and spectroscopy, astronomers can derive the dynamical masses of the individual components, as well as their surface temperatures, radii and luminosities.

Eclipsing binary systems have been used as "standard candles" to measure the distances, in particular in systems where the individual stars' properties were well known. These systems are also of paramount importance to test and calibrate stellar evolution models. The accurate determination of the individual component properties, obtained from these systems, provides a direct comparison between theory and observation. 

In systems where the inclination is low and do not show eclipses, the light curves can show variations due to the deformation and the changing apparent surface area of the stars. These systems are called {\it ellipsoidal variables}. Unlike eclipsing binaries, ellipsoidal variables do not show sharp drops in brightness but rather smooth, sinusoidal light curves due to the continuous change in the visible surface area. In these systems, the stars are close enough that their mutual gravitational attraction causes them to become tidally distorted, taking on an ellipsoidal (elongated) shape rather than a perfect sphere. The distortion is most significant for the stars that are filling or nearly filling their Roche lobe.

\subsection{Interferometric and astrometric binaries}\label{sec4:subsec3}
Interferometry and astrometry provide positions with time that allow us to track the orbital motion of the companions, directly or indirectly. Interferometry is an observational technique that aims to combine the light collected from multiple telescopes in order to create the resolving power of a much larger telescope. It uses interference patterns (constructive or destructive) that contain information about the source's structure. Interferometric observations of binary systems allow us to detect the relative motion of the companion star with respect to the primary object, reconstruct the orbit, and derive the orbital parameters, such as the orbital period, eccentricity, semi-major axis, total mass of the system. When combined with the spectroscopic parameters, the astronomer is able to measure the exact masses of the individual components. \\ 
 
Astrometric binaries are binary systems in which only one star is directly detected, but the presence of an invisible companion star is inferred through its gravitational influence on the visible star's motion. Its presence causes the visible star to wobble slightly around the common center of mass. The detection of this wobble through precise measurements of the star's position (including the linear change due to the proper motion) may allow us to detect invisible companions like black holes or planets or very faint ones like white or brown dwarfs, or dim stars. The space mission Gaia has revolutionized the detection of astrometric binaries by providing extremely precise measurements of stellar positions. The size and period of the wobble, combined with the distance of the star from Earth, can be used to estimate the mass of the companion. Astrometric measurements can provide information about the orbit of the binary system, including the orbital period, the eccentricity, and the orientation in space, as well as on the masses of the components. 
%\begin{figure}[b]
%\centering
%\includegraphics[width=0.65\textwidth]{blankfig}
%\caption{A conservation relationship can also be written for electric charge, but %in mesoscale modeling, electromagnetic effects are not considered to be %dynamically or thermodynamically important on the model-resolved mesoscale.}
%\label{chap1:fig2}
%\end{figure}

\section{Conclusions}

Fundamental observational properties of stars serve as the foundation for understanding stellar physics and evolution. They provide essential insights into the life cycles, structure, and ultimate fates of stars. Understanding these observables is the first step toward a deeper exploration of the stellar physics. Indeed, the basic stellar observables presented in this Chapter only represent the tip of the iceberg and many more concepts about stellar astrophysics are crucial to have a global view of stellar evolution. With increasing observational capabilities and improved measurements of fundamental stellar parameters, the comparison between basic stellar observables and theoretical stellar evolution models is become critical for advancing our knowledge of stellar physics. 

\begin{ack}[Acknowledgments]

I would like to thank Eric Gosset, Patricia Lampens and Thibault Merle for discussions and feedbacks on the manuscript. I also acknowledges the Belgian Science Policy Office (BELSPO) for the financial support. 

\end{ack}

\seealso{\citet{kopal59}; \citet{serenelli21}; \citet{marchant23}}

\bibliographystyle{Harvard}
\bibliography{observable}

\begin{thebibliography*}{48}
\providecommand{\bibtype}[1]{}
\providecommand{\natexlab}[1]{#1}
{\catcode`\|=0\catcode`\#=12\catcode`\@=11\catcode`\\=12
|immediate|write|@auxout{\expandafter\ifx\csname natexlab\endcsname\relax\gdef\natexlab#1{#1}\fi}}
\renewcommand{\url}[1]{{\tt #1}}
\providecommand{\urlprefix}{URL }
\expandafter\ifx\csname urlstyle\endcsname\relax
  \providecommand{\doi}[1]{doi:\discretionary{}{}{}#1}\else
  \providecommand{\doi}{doi:\discretionary{}{}{}\begingroup \urlstyle{rm}\Url}\fi
\providecommand{\bibinfo}[2]{#2}
\providecommand{\eprint}[2][]{\url{#2}}

\bibtype{Article}%
\bibitem[{Bergemann} et al.(2016)]{Bergemann16}
\bibinfo{author}{{Bergemann} M}, \bibinfo{author}{{Serenelli} A}, \bibinfo{author}{{Sch{\"o}nrich} R}, \bibinfo{author}{{Ruchti} G}, \bibinfo{author}{{Korn} A}, \bibinfo{author}{{Hekker} S}, \bibinfo{author}{{Kovalev} M}, \bibinfo{author}{{Mashonkina} L}, \bibinfo{author}{{Gilmore} G}, \bibinfo{author}{{Randich} S}, \bibinfo{author}{{Asplund} M}, \bibinfo{author}{{Rix} HW}, \bibinfo{author}{{Casey} AR}, \bibinfo{author}{{Jofre} P}, \bibinfo{author}{{Pancino} E}, \bibinfo{author}{{Recio-Blanco} A}, \bibinfo{author}{{de Laverny} P}, \bibinfo{author}{{Smiljanic} R}, \bibinfo{author}{{Tautvaisiene} G}, \bibinfo{author}{{Bayo} A}, \bibinfo{author}{{Lewis} J}, \bibinfo{author}{{Koposov} S}, \bibinfo{author}{{Hourihane} A}, \bibinfo{author}{{Worley} C}, \bibinfo{author}{{Morbidelli} L}, \bibinfo{author}{{Franciosini} E}, \bibinfo{author}{{Sacco} G}, \bibinfo{author}{{Magrini} L}, \bibinfo{author}{{Damiani} F} and  \bibinfo{author}{{Bestenlehner} JM} (\bibinfo{year}{2016}), \bibinfo{month}{Oct.}
\bibinfo{title}{{The Gaia-ESO Survey: Hydrogen lines in red giants directly trace stellar mass}}.
\bibinfo{journal}{{\em A\&A}} \bibinfo{volume}{594}, \bibinfo{eid}{A120}. \bibinfo{doi}{\doi{10.1051/0004-6361/201528010}}.
\eprint{1606.05661}.

\bibtype{Article}%
\bibitem[{Carroll}(1933)]{carroll33}
\bibinfo{author}{{Carroll} JA} (\bibinfo{year}{1933}), \bibinfo{month}{May}.
\bibinfo{title}{{The spectroscopic determination of stellar rotation and its effect on line profiles}}.
\bibinfo{journal}{{\em MNRAS}} \bibinfo{volume}{93}: \bibinfo{pages}{478--507}. \bibinfo{doi}{\doi{10.1093/mnras/93.7.478}}.

\bibtype{Inproceedings}%
\bibitem[{Castelli} and {Kurucz}(2003)]{castelli03}
\bibinfo{author}{{Castelli} F} and  \bibinfo{author}{{Kurucz} RL} (\bibinfo{year}{2003}), \bibinfo{month}{Jan.}, \bibinfo{title}{{New Grids of ATLAS9 Model Atmospheres}}, \bibinfo{editor}{{Piskunov} N}, \bibinfo{editor}{{Weiss} WW} and  \bibinfo{editor}{{Gray} DF}, (Eds.), \bibinfo{booktitle}{Modelling of Stellar Atmospheres}, \bibinfo{series}{IAU Symposium}, \bibinfo{volume}{210}, pp. \bibinfo{pages}{A20}, \eprint{astro-ph/0405087}.

\bibtype{Article}%
\bibitem[{Conti}(1975)]{conti76}
\bibinfo{author}{{Conti} PS} (\bibinfo{year}{1975}), \bibinfo{month}{Jan.}
\bibinfo{title}{{On the relationship between Of and WR stars.}}
\bibinfo{journal}{{\em Memoires of the Societe Royale des Sciences de Liege}} \bibinfo{volume}{9}: \bibinfo{pages}{193--212}.

\bibtype{Article}%
\bibitem[{Crowther}(2007)]{crowther07}
\bibinfo{author}{{Crowther} PA} (\bibinfo{year}{2007}), \bibinfo{month}{Sep.}
\bibinfo{title}{{Physical Properties of Wolf-Rayet Stars}}.
\bibinfo{journal}{{\em ARA\&A}} \bibinfo{volume}{45} (\bibinfo{number}{1}): \bibinfo{pages}{177--219}. \bibinfo{doi}{\doi{10.1146/annurev.astro.45.051806.110615}}.
\eprint{astro-ph/0610356}.

\bibtype{Article}%
\bibitem[{Do Nascimento} et al.(2009)]{DoNascimento09}
\bibinfo{author}{{Do Nascimento} J.~D. J}, \bibinfo{author}{{Castro} M}, \bibinfo{author}{{Mel{\'e}ndez} J}, \bibinfo{author}{{Bazot} M}, \bibinfo{author}{{Th{\'e}ado} S}, \bibinfo{author}{{Porto de Mello} GF} and  \bibinfo{author}{{de Medeiros} JR} (\bibinfo{year}{2009}), \bibinfo{month}{Jul.}
\bibinfo{title}{{Age and mass of solar twins constrained by lithium abundance}}.
\bibinfo{journal}{{\em A\&A}} \bibinfo{volume}{501} (\bibinfo{number}{2}): \bibinfo{pages}{687--694}. \bibinfo{doi}{\doi{10.1051/0004-6361/200911935}}.
\eprint{0904.3580}.

\bibtype{Article}%
\bibitem[{Eggleton}(1983)]{Eggleton83}
\bibinfo{author}{{Eggleton} PP} (\bibinfo{year}{1983}), \bibinfo{month}{May}.
\bibinfo{title}{{Aproximations to the radii of Roche lobes.}}
\bibinfo{journal}{{\em ApJ}} \bibinfo{volume}{268}: \bibinfo{pages}{368--369}. \bibinfo{doi}{\doi{10.1086/160960}}.

\bibtype{Article}%
\bibitem[{Eker} et al.(2018)]{eker18}
\bibinfo{author}{{Eker} Z}, \bibinfo{author}{{Bak{\i}{\c{s}}} V}, \bibinfo{author}{{Bilir} S}, \bibinfo{author}{{Soydugan} F}, \bibinfo{author}{{Steer} I}, \bibinfo{author}{{Soydugan} E}, \bibinfo{author}{{Bak{\i}{\c{s}}} H}, \bibinfo{author}{{Ali{\c{c}}avu{\c{s}}} F}, \bibinfo{author}{{Aslan} G} and  \bibinfo{author}{{Alpsoy} M} (\bibinfo{year}{2018}), \bibinfo{month}{Oct.}
\bibinfo{title}{{Interrelated main-sequence mass-luminosity, mass-radius, and mass-effective temperature relations}}.
\bibinfo{journal}{{\em MNRAS}} \bibinfo{volume}{479} (\bibinfo{number}{4}): \bibinfo{pages}{5491--5511}. \bibinfo{doi}{\doi{10.1093/mnras/sty1834}}.
\eprint{1807.02568}.

\bibtype{Article}%
\bibitem[{Gaia Collaboration} et al.(2018)]{babusiaux18}
\bibinfo{author}{{Gaia Collaboration}}, \bibinfo{author}{{Babusiaux} C}, \bibinfo{author}{{van Leeuwen} F}, \bibinfo{author}{{Barstow} MA}, \bibinfo{author}{{Jordi} C}, \bibinfo{author}{{Vallenari} A}, \bibinfo{author}{{Bossini} D}, \bibinfo{author}{{Bressan} A}, \bibinfo{author}{{Cantat-Gaudin} T}, \bibinfo{author}{{van Leeuwen} M}, \bibinfo{author}{{Brown} AGA}, \bibinfo{author}{{Prusti} T}, \bibinfo{author}{{de Bruijne} JHJ}, \bibinfo{author}{{Bailer-Jones} CAL}, \bibinfo{author}{{Biermann} M}, \bibinfo{author}{{Evans} DW}, \bibinfo{author}{{Eyer} L}, \bibinfo{author}{{Jansen} F}, \bibinfo{author}{{Klioner} SA}, \bibinfo{author}{{Lammers} U}, \bibinfo{author}{{Lindegren} L}, \bibinfo{author}{{Luri} X}, \bibinfo{author}{{Mignard} F}, \bibinfo{author}{{Panem} C}, \bibinfo{author}{{Pourbaix} D}, \bibinfo{author}{{Randich} S}, \bibinfo{author}{{Sartoretti} P}, \bibinfo{author}{{Siddiqui} HI}, \bibinfo{author}{{Soubiran} C}, \bibinfo{author}{{Walton} NA}, \bibinfo{author}{{Arenou} F}, \bibinfo{author}{{Bastian}
  U}, \bibinfo{author}{{Cropper} M}, \bibinfo{author}{{Drimmel} R}, \bibinfo{author}{{Katz} D}, \bibinfo{author}{{Lattanzi} MG}, \bibinfo{author}{{Bakker} J}, \bibinfo{author}{{Cacciari} C}, \bibinfo{author}{{Casta{\~n}eda} J}, \bibinfo{author}{{Chaoul} L}, \bibinfo{author}{{Cheek} N}, \bibinfo{author}{{De Angeli} F}, \bibinfo{author}{{Fabricius} C}, \bibinfo{author}{{Guerra} R}, \bibinfo{author}{{Holl} B}, \bibinfo{author}{{Masana} E}, \bibinfo{author}{{Messineo} R}, \bibinfo{author}{{Mowlavi} N}, \bibinfo{author}{{Nienartowicz} K}, \bibinfo{author}{{Panuzzo} P}, \bibinfo{author}{{Portell} J}, \bibinfo{author}{{Riello} M}, \bibinfo{author}{{Seabroke} GM}, \bibinfo{author}{{Tanga} P}, \bibinfo{author}{{Th{\'e}venin} F}, \bibinfo{author}{{Gracia-Abril} G}, \bibinfo{author}{{Comoretto} G}, \bibinfo{author}{{Garcia-Reinaldos} M}, \bibinfo{author}{{Teyssier} D}, \bibinfo{author}{{Altmann} M}, \bibinfo{author}{{Andrae} R}, \bibinfo{author}{{Audard} M}, \bibinfo{author}{{Bellas-Velidis} I},
  \bibinfo{author}{{Benson} K}, \bibinfo{author}{{Berthier} J}, \bibinfo{author}{{Blomme} R}, \bibinfo{author}{{Burgess} P}, \bibinfo{author}{{Busso} G}, \bibinfo{author}{{Carry} B}, \bibinfo{author}{{Cellino} A}, \bibinfo{author}{{Clementini} G}, \bibinfo{author}{{Clotet} M}, \bibinfo{author}{{Creevey} O}, \bibinfo{author}{{Davidson} M}, \bibinfo{author}{{De Ridder} J}, \bibinfo{author}{{Delchambre} L}, \bibinfo{author}{{Dell'Oro} A}, \bibinfo{author}{{Ducourant} C}, \bibinfo{author}{{Fern{\'a}ndez-Hern{\'a}ndez} J}, \bibinfo{author}{{Fouesneau} M}, \bibinfo{author}{{Fr{\'e}mat} Y}, \bibinfo{author}{{Galluccio} L}, \bibinfo{author}{{Garc{\'\i}a-Torres} M}, \bibinfo{author}{{Gonz{\'a}lez-N{\'u}{\~n}ez} J}, \bibinfo{author}{{Gonz{\'a}lez-Vidal} JJ}, \bibinfo{author}{{Gosset} E}, \bibinfo{author}{{Guy} LP}, \bibinfo{author}{{Halbwachs} JL}, \bibinfo{author}{{Hambly} NC}, \bibinfo{author}{{Harrison} DL}, \bibinfo{author}{{Hern{\'a}ndez} J}, \bibinfo{author}{{Hestroffer} D}, \bibinfo{author}{{Hodgkin} ST},
  \bibinfo{author}{{Hutton} A}, \bibinfo{author}{{Jasniewicz} G}, \bibinfo{author}{{Jean-Antoine-Piccolo} A}, \bibinfo{author}{{Jordan} S}, \bibinfo{author}{{Korn} AJ}, \bibinfo{author}{{Krone-Martins} A}, \bibinfo{author}{{Lanzafame} AC}, \bibinfo{author}{{Lebzelter} T}, \bibinfo{author}{{L{\"o}ffler} W}, \bibinfo{author}{{Manteiga} M}, \bibinfo{author}{{Marrese} PM}, \bibinfo{author}{{Mart{\'\i}n-Fleitas} JM}, \bibinfo{author}{{Moitinho} A}, \bibinfo{author}{{Mora} A}, \bibinfo{author}{{Muinonen} K}, \bibinfo{author}{{Osinde} J}, \bibinfo{author}{{Pancino} E}, \bibinfo{author}{{Pauwels} T}, \bibinfo{author}{{Petit} JM}, \bibinfo{author}{{Recio-Blanco} A}, \bibinfo{author}{{Richards} PJ}, \bibinfo{author}{{Rimoldini} L}, \bibinfo{author}{{Robin} AC}, \bibinfo{author}{{Sarro} LM}, \bibinfo{author}{{Siopis} C}, \bibinfo{author}{{Smith} M}, \bibinfo{author}{{Sozzetti} A}, \bibinfo{author}{{S{\"u}veges} M}, \bibinfo{author}{{Torra} J}, \bibinfo{author}{{van Reeven} W}, \bibinfo{author}{{Abbas} U},
  \bibinfo{author}{{Abreu Aramburu} A}, \bibinfo{author}{{Accart} S}, \bibinfo{author}{{Aerts} C}, \bibinfo{author}{{Altavilla} G}, \bibinfo{author}{{{\'A}lvarez} MA}, \bibinfo{author}{{Alvarez} R}, \bibinfo{author}{{Alves} J}, \bibinfo{author}{{Anderson} RI}, \bibinfo{author}{{Andrei} AH}, \bibinfo{author}{{Anglada Varela} E}, \bibinfo{author}{{Antiche} E}, \bibinfo{author}{{Antoja} T}, \bibinfo{author}{{Arcay} B}, \bibinfo{author}{{Astraatmadja} TL}, \bibinfo{author}{{Bach} N}, \bibinfo{author}{{Baker} SG}, \bibinfo{author}{{Balaguer-N{\'u}{\~n}ez} L}, \bibinfo{author}{{Balm} P}, \bibinfo{author}{{Barache} C}, \bibinfo{author}{{Barata} C}, \bibinfo{author}{{Barbato} D}, \bibinfo{author}{{Barblan} F}, \bibinfo{author}{{Barklem} PS}, \bibinfo{author}{{Barrado} D}, \bibinfo{author}{{Barros} M}, \bibinfo{author}{{Bartholom{\'e} Mu{\~n}oz} L}, \bibinfo{author}{{Bassilana} JL}, \bibinfo{author}{{Becciani} U}, \bibinfo{author}{{Bellazzini} M}, \bibinfo{author}{{Berihuete} A}, \bibinfo{author}{{Bertone} S},
  \bibinfo{author}{{Bianchi} L}, \bibinfo{author}{{Bienaym{\'e}} O}, \bibinfo{author}{{Blanco-Cuaresma} S}, \bibinfo{author}{{Boch} T}, \bibinfo{author}{{Boeche} C}, \bibinfo{author}{{Bombrun} A}, \bibinfo{author}{{Borrachero} R}, \bibinfo{author}{{Bouquillon} S}, \bibinfo{author}{{Bourda} G}, \bibinfo{author}{{Bragaglia} A}, \bibinfo{author}{{Bramante} L}, \bibinfo{author}{{Breddels} MA}, \bibinfo{author}{{Brouillet} N}, \bibinfo{author}{{Br{\"u}semeister} T}, \bibinfo{author}{{Brugaletta} E}, \bibinfo{author}{{Bucciarelli} B}, \bibinfo{author}{{Burlacu} A}, \bibinfo{author}{{Busonero} D}, \bibinfo{author}{{Butkevich} AG}, \bibinfo{author}{{Buzzi} R}, \bibinfo{author}{{Caffau} E}, \bibinfo{author}{{Cancelliere} R}, \bibinfo{author}{{Cannizzaro} G}, \bibinfo{author}{{Carballo} R}, \bibinfo{author}{{Carlucci} T}, \bibinfo{author}{{Carrasco} JM}, \bibinfo{author}{{Casamiquela} L}, \bibinfo{author}{{Castellani} M}, \bibinfo{author}{{Castro-Ginard} A}, \bibinfo{author}{{Charlot} P}, \bibinfo{author}{{Chemin} L},
  \bibinfo{author}{{Chiavassa} A}, \bibinfo{author}{{Cocozza} G}, \bibinfo{author}{{Costigan} G}, \bibinfo{author}{{Cowell} S}, \bibinfo{author}{{Crifo} F}, \bibinfo{author}{{Crosta} M}, \bibinfo{author}{{Crowley} C}, \bibinfo{author}{{Cuypers} J}, \bibinfo{author}{{Dafonte} C}, \bibinfo{author}{{Damerdji} Y}, \bibinfo{author}{{Dapergolas} A}, \bibinfo{author}{{David} P}, \bibinfo{author}{{David} M}, \bibinfo{author}{{de Laverny} P}, \bibinfo{author}{{De Luise} F}, \bibinfo{author}{{De March} R}, \bibinfo{author}{{de Martino} D}, \bibinfo{author}{{de Souza} R}, \bibinfo{author}{{de Torres} A}, \bibinfo{author}{{Debosscher} J}, \bibinfo{author}{{del Pozo} E}, \bibinfo{author}{{Delbo} M}, \bibinfo{author}{{Delgado} A}, \bibinfo{author}{{Delgado} HE}, \bibinfo{author}{{Diakite} S}, \bibinfo{author}{{Diener} C}, \bibinfo{author}{{Distefano} E}, \bibinfo{author}{{Dolding} C}, \bibinfo{author}{{Drazinos} P}, \bibinfo{author}{{Dur{\'a}n} J}, \bibinfo{author}{{Edvardsson} B}, \bibinfo{author}{{Enke} H},
  \bibinfo{author}{{Eriksson} K}, \bibinfo{author}{{Esquej} P}, \bibinfo{author}{{Eynard Bontemps} G}, \bibinfo{author}{{Fabre} C}, \bibinfo{author}{{Fabrizio} M}, \bibinfo{author}{{Faigler} S}, \bibinfo{author}{{Falc{\~a}o} AJ}, \bibinfo{author}{{Farr{\`a}s Casas} M}, \bibinfo{author}{{Federici} L}, \bibinfo{author}{{Fedorets} G}, \bibinfo{author}{{Fernique} P}, \bibinfo{author}{{Figueras} F}, \bibinfo{author}{{Filippi} F}, \bibinfo{author}{{Findeisen} K}, \bibinfo{author}{{Fonti} A}, \bibinfo{author}{{Fraile} E}, \bibinfo{author}{{Fraser} M}, \bibinfo{author}{{Fr{\'e}zouls} B}, \bibinfo{author}{{Gai} M}, \bibinfo{author}{{Galleti} S}, \bibinfo{author}{{Garabato} D}, \bibinfo{author}{{Garc{\'\i}a-Sedano} F}, \bibinfo{author}{{Garofalo} A}, \bibinfo{author}{{Garralda} N}, \bibinfo{author}{{Gavel} A}, \bibinfo{author}{{Gavras} P}, \bibinfo{author}{{Gerssen} J}, \bibinfo{author}{{Geyer} R}, \bibinfo{author}{{Giacobbe} P}, \bibinfo{author}{{Gilmore} G}, \bibinfo{author}{{Girona} S}, \bibinfo{author}{{Giuffrida}
  G}, \bibinfo{author}{{Glass} F}, \bibinfo{author}{{Gomes} M}, \bibinfo{author}{{Granvik} M}, \bibinfo{author}{{Gueguen} A}, \bibinfo{author}{{Guerrier} A}, \bibinfo{author}{{Guiraud} J}, \bibinfo{author}{{Guti{\'e}} R}, \bibinfo{author}{{Haigron} R}, \bibinfo{author}{{Hatzidimitriou} D}, \bibinfo{author}{{Hauser} M}, \bibinfo{author}{{Haywood} M}, \bibinfo{author}{{Heiter} U}, \bibinfo{author}{{Helmi} A}, \bibinfo{author}{{Heu} J}, \bibinfo{author}{{Hilger} T}, \bibinfo{author}{{Hobbs} D}, \bibinfo{author}{{Hofmann} W}, \bibinfo{author}{{Holland} G}, \bibinfo{author}{{Huckle} HE}, \bibinfo{author}{{Hypki} A}, \bibinfo{author}{{Icardi} V}, \bibinfo{author}{{Jan{\ss}en} K}, \bibinfo{author}{{Jevardat de Fombelle} G}, \bibinfo{author}{{Jonker} PG}, \bibinfo{author}{{Juh{\'a}sz} {\'A}L}, \bibinfo{author}{{Julbe} F}, \bibinfo{author}{{Karampelas} A}, \bibinfo{author}{{Kewley} A}, \bibinfo{author}{{Klar} J}, \bibinfo{author}{{Kochoska} A}, \bibinfo{author}{{Kohley} R}, \bibinfo{author}{{Kolenberg} K},
  \bibinfo{author}{{Kontizas} M}, \bibinfo{author}{{Kontizas} E}, \bibinfo{author}{{Koposov} SE}, \bibinfo{author}{{Kordopatis} G}, \bibinfo{author}{{Kostrzewa-Rutkowska} Z}, \bibinfo{author}{{Koubsky} P}, \bibinfo{author}{{Lambert} S}, \bibinfo{author}{{Lanza} AF}, \bibinfo{author}{{Lasne} Y}, \bibinfo{author}{{Lavigne} JB}, \bibinfo{author}{{Le Fustec} Y}, \bibinfo{author}{{Le Poncin-Lafitte} C}, \bibinfo{author}{{Lebreton} Y}, \bibinfo{author}{{Leccia} S}, \bibinfo{author}{{Leclerc} N}, \bibinfo{author}{{Lecoeur-Taibi} I}, \bibinfo{author}{{Lenhardt} H}, \bibinfo{author}{{Leroux} F}, \bibinfo{author}{{Liao} S}, \bibinfo{author}{{Licata} E}, \bibinfo{author}{{Lindstr{\o}m} HEP}, \bibinfo{author}{{Lister} TA}, \bibinfo{author}{{Livanou} E}, \bibinfo{author}{{Lobel} A}, \bibinfo{author}{{L{\'o}pez} M}, \bibinfo{author}{{Managau} S}, \bibinfo{author}{{Mann} RG}, \bibinfo{author}{{Mantelet} G}, \bibinfo{author}{{Marchal} O}, \bibinfo{author}{{Marchant} JM}, \bibinfo{author}{{Marconi} M},
  \bibinfo{author}{{Marinoni} S}, \bibinfo{author}{{Marschalk{\'o}} G}, \bibinfo{author}{{Marshall} DJ}, \bibinfo{author}{{Martino} M}, \bibinfo{author}{{Marton} G}, \bibinfo{author}{{Mary} N}, \bibinfo{author}{{Massari} D}, \bibinfo{author}{{Matijevi{\v{c}}} G}, \bibinfo{author}{{Mazeh} T}, \bibinfo{author}{{McMillan} PJ}, \bibinfo{author}{{Messina} S}, \bibinfo{author}{{Michalik} D}, \bibinfo{author}{{Millar} NR}, \bibinfo{author}{{Molina} D}, \bibinfo{author}{{Molinaro} R}, \bibinfo{author}{{Moln{\'a}r} L}, \bibinfo{author}{{Montegriffo} P}, \bibinfo{author}{{Mor} R}, \bibinfo{author}{{Morbidelli} R}, \bibinfo{author}{{Morel} T}, \bibinfo{author}{{Morris} D}, \bibinfo{author}{{Mulone} AF}, \bibinfo{author}{{Muraveva} T}, \bibinfo{author}{{Musella} I}, \bibinfo{author}{{Nelemans} G}, \bibinfo{author}{{Nicastro} L}, \bibinfo{author}{{Noval} L}, \bibinfo{author}{{O'Mullane} W}, \bibinfo{author}{{Ord{\'e}novic} C}, \bibinfo{author}{{Ord{\'o}{\~n}ez-Blanco} D}, \bibinfo{author}{{Osborne} P},
  \bibinfo{author}{{Pagani} C}, \bibinfo{author}{{Pagano} I}, \bibinfo{author}{{Pailler} F}, \bibinfo{author}{{Palacin} H}, \bibinfo{author}{{Palaversa} L}, \bibinfo{author}{{Panahi} A}, \bibinfo{author}{{Pawlak} M}, \bibinfo{author}{{Piersimoni} AM}, \bibinfo{author}{{Pineau} FX}, \bibinfo{author}{{Plachy} E}, \bibinfo{author}{{Plum} G}, \bibinfo{author}{{Poggio} E}, \bibinfo{author}{{Poujoulet} E}, \bibinfo{author}{{Pr{\v{s}}a} A}, \bibinfo{author}{{Pulone} L}, \bibinfo{author}{{Racero} E}, \bibinfo{author}{{Ragaini} S}, \bibinfo{author}{{Rambaux} N}, \bibinfo{author}{{Ramos-Lerate} M}, \bibinfo{author}{{Regibo} S}, \bibinfo{author}{{Reyl{\'e}} C}, \bibinfo{author}{{Riclet} F}, \bibinfo{author}{{Ripepi} V}, \bibinfo{author}{{Riva} A}, \bibinfo{author}{{Rivard} A}, \bibinfo{author}{{Rixon} G}, \bibinfo{author}{{Roegiers} T}, \bibinfo{author}{{Roelens} M}, \bibinfo{author}{{Romero-G{\'o}mez} M}, \bibinfo{author}{{Rowell} N}, \bibinfo{author}{{Royer} F}, \bibinfo{author}{{Ruiz-Dern} L},
  \bibinfo{author}{{Sadowski} G}, \bibinfo{author}{{Sagrist{\`a} Sell{\'e}s} T}, \bibinfo{author}{{Sahlmann} J}, \bibinfo{author}{{Salgado} J}, \bibinfo{author}{{Salguero} E}, \bibinfo{author}{{Sanna} N}, \bibinfo{author}{{Santana-Ros} T}, \bibinfo{author}{{Sarasso} M}, \bibinfo{author}{{Savietto} H}, \bibinfo{author}{{Schultheis} M}, \bibinfo{author}{{Sciacca} E}, \bibinfo{author}{{Segol} M}, \bibinfo{author}{{Segovia} JC}, \bibinfo{author}{{S{\'e}gransan} D}, \bibinfo{author}{{Shih} IC}, \bibinfo{author}{{Siltala} L}, \bibinfo{author}{{Silva} AF}, \bibinfo{author}{{Smart} RL}, \bibinfo{author}{{Smith} KW}, \bibinfo{author}{{Solano} E}, \bibinfo{author}{{Solitro} F}, \bibinfo{author}{{Sordo} R}, \bibinfo{author}{{Soria Nieto} S}, \bibinfo{author}{{Souchay} J}, \bibinfo{author}{{Spagna} A}, \bibinfo{author}{{Spoto} F}, \bibinfo{author}{{Stampa} U}, \bibinfo{author}{{Steele} IA}, \bibinfo{author}{{Steidelm{\"u}ller} H}, \bibinfo{author}{{Stephenson} CA}, \bibinfo{author}{{Stoev} H}, \bibinfo{author}{{Suess}
  FF}, \bibinfo{author}{{Surdej} J}, \bibinfo{author}{{Szabados} L}, \bibinfo{author}{{Szegedi-Elek} E}, \bibinfo{author}{{Tapiador} D}, \bibinfo{author}{{Taris} F}, \bibinfo{author}{{Tauran} G}, \bibinfo{author}{{Taylor} MB}, \bibinfo{author}{{Teixeira} R}, \bibinfo{author}{{Terrett} D}, \bibinfo{author}{{Teyssandier} P}, \bibinfo{author}{{Thuillot} W}, \bibinfo{author}{{Titarenko} A}, \bibinfo{author}{{Torra Clotet} F}, \bibinfo{author}{{Turon} C}, \bibinfo{author}{{Ulla} A}, \bibinfo{author}{{Utrilla} E}, \bibinfo{author}{{Uzzi} S}, \bibinfo{author}{{Vaillant} M}, \bibinfo{author}{{Valentini} G}, \bibinfo{author}{{Valette} V}, \bibinfo{author}{{van Elteren} A}, \bibinfo{author}{{Van Hemelryck} E}, \bibinfo{author}{{Vaschetto} M}, \bibinfo{author}{{Vecchiato} A}, \bibinfo{author}{{Veljanoski} J}, \bibinfo{author}{{Viala} Y}, \bibinfo{author}{{Vicente} D}, \bibinfo{author}{{Vogt} S}, \bibinfo{author}{{von Essen} C}, \bibinfo{author}{{Voss} H}, \bibinfo{author}{{Votruba} V}, \bibinfo{author}{{Voutsinas} S},
  \bibinfo{author}{{Walmsley} G}, \bibinfo{author}{{Weiler} M}, \bibinfo{author}{{Wertz} O}, \bibinfo{author}{{Wevers} T}, \bibinfo{author}{{Wyrzykowski} {\L}}, \bibinfo{author}{{Yoldas} A}, \bibinfo{author}{{{\v{Z}}erjal} M}, \bibinfo{author}{{Ziaeepour} H}, \bibinfo{author}{{Zorec} J}, \bibinfo{author}{{Zschocke} S}, \bibinfo{author}{{Zucker} S}, \bibinfo{author}{{Zurbach} C} and  \bibinfo{author}{{Zwitter} T} (\bibinfo{year}{2018}), \bibinfo{month}{Aug.}
\bibinfo{title}{{Gaia Data Release 2. Observational Hertzsprung-Russell diagrams}}.
\bibinfo{journal}{{\em A\&A}} \bibinfo{volume}{616}, \bibinfo{eid}{A10}. \bibinfo{doi}{\doi{10.1051/0004-6361/201832843}}.
\eprint{1804.09378}.

\bibtype{Book}%
\bibitem[{Gray}(1976)]{gray76}
\bibinfo{author}{{Gray} DF} (\bibinfo{year}{1976}).
\bibinfo{title}{{The observation and analysis of stellar photospheres}}.

\bibtype{Book}%
\bibitem[{Gray}(2005)]{gray05}
\bibinfo{author}{{Gray} DF} (\bibinfo{year}{2005}).
\bibinfo{title}{{The Observation and Analysis of Stellar Photospheres}}.
\bibinfo{doi}{\doi{10.1017/CBO9781316036570}}.

\bibtype{Article}%
\bibitem[{Groenewegen}(2021)]{groenewegen21}
\bibinfo{author}{{Groenewegen} MAT} (\bibinfo{year}{2021}), \bibinfo{month}{Oct.}
\bibinfo{title}{{The parallax zero-point offset from Gaia EDR3 data}}.
\bibinfo{journal}{{\em A\&A}} \bibinfo{volume}{654}, \bibinfo{eid}{A20}. \bibinfo{doi}{\doi{10.1051/0004-6361/202140862}}.
\eprint{2106.08128}.

\bibtype{Inproceedings}%
\bibitem[{Groenewegen}(2024)]{groenewegen24}
\bibinfo{author}{{Groenewegen} MAT} (\bibinfo{year}{2024}), \bibinfo{month}{Jan.}, \bibinfo{title}{{Primary Period-Luminosity-Relation Calibrators in the Milky Way: Cepheids and RR Lyrae Physical basis, Calibration, and Applications}}, \bibinfo{editor}{{de Grijs} R}, \bibinfo{editor}{{Whitelock} PA} and  \bibinfo{editor}{{Catelan} M}, (Eds.), \bibinfo{booktitle}{IAU Symposium}, \bibinfo{series}{IAU Symposium}, \bibinfo{volume}{376},  \bibinfo{pages}{128--149}, \eprint{2307.03033}.

\bibtype{Article}%
\bibitem[{Gustafsson} et al.(1975)]{gustafsson75}
\bibinfo{author}{{Gustafsson} B}, \bibinfo{author}{{Bell} RA}, \bibinfo{author}{{Eriksson} K} and  \bibinfo{author}{{Nordlund} A} (\bibinfo{year}{1975}), \bibinfo{month}{Sep.}
\bibinfo{title}{{A grid of model atmospheres for metal-deficient giant stars. I.}}
\bibinfo{journal}{{\em A\&A}} \bibinfo{volume}{42}: \bibinfo{pages}{407--432}.

\bibtype{Article}%
\bibitem[{Gustafsson} et al.(2008)]{gustafsson08}
\bibinfo{author}{{Gustafsson} B}, \bibinfo{author}{{Edvardsson} B}, \bibinfo{author}{{Eriksson} K}, \bibinfo{author}{{J{\o}rgensen} UG}, \bibinfo{author}{{Nordlund} {\r{A}}} and  \bibinfo{author}{{Plez} B} (\bibinfo{year}{2008}), \bibinfo{month}{Aug.}
\bibinfo{title}{{A grid of MARCS model atmospheres for late-type stars. I. Methods and general properties}}.
\bibinfo{journal}{{\em A\&A}} \bibinfo{volume}{486} (\bibinfo{number}{3}): \bibinfo{pages}{951--970}. \bibinfo{doi}{\doi{10.1051/0004-6361:200809724}}.
\eprint{0805.0554}.

\bibtype{Article}%
\bibitem[{Hamann} and {Gr{\"a}fener}(2003)]{hamann03}
\bibinfo{author}{{Hamann} WR} and  \bibinfo{author}{{Gr{\"a}fener} G} (\bibinfo{year}{2003}), \bibinfo{month}{Nov.}
\bibinfo{title}{{A temperature correction method for expanding atmospheres}}.
\bibinfo{journal}{{\em A\&A}} \bibinfo{volume}{410}: \bibinfo{pages}{993--1000}. \bibinfo{doi}{\doi{10.1051/0004-6361:20031308}}.

\bibtype{Article}%
\bibitem[{Hauschildt} et al.(1999)]{hauschildt99}
\bibinfo{author}{{Hauschildt} PH}, \bibinfo{author}{{Allard} F} and  \bibinfo{author}{{Baron} E} (\bibinfo{year}{1999}), \bibinfo{month}{Feb.}
\bibinfo{title}{{The NextGen Model Atmosphere Grid for 3000<=T$_{eff}$<=10,000 K}}.
\bibinfo{journal}{{\em ApJ}} \bibinfo{volume}{512} (\bibinfo{number}{1}): \bibinfo{pages}{377--385}. \bibinfo{doi}{\doi{10.1086/306745}}.
\eprint{astro-ph/9807286}.

\bibtype{incollection}%
\bibitem[{Herrero} et al.(1992)]{herrero92}
\bibinfo{author}{{Herrero} A}, \bibinfo{author}{{Kudritzki} RP}, \bibinfo{author}{{Vilchez} JM}, \bibinfo{author}{{Kunze} D}, \bibinfo{author}{{Butler} K} and  \bibinfo{author}{{Haser} S} (\bibinfo{year}{1992}), \bibinfo{title}{{The mass and helium discrepancy in massive young stars}}, \bibinfo{editor}{{Heber} U} and  \bibinfo{editor}{{Jeffery} CS}, (Eds.), \bibinfo{booktitle}{The Atmospheres of Early-Type Stars}, \bibinfo{volume}{401}, pp.~\bibinfo{pages}{21}.

\bibtype{Article}%
\bibitem[{Hillier} and {Miller}(1998)]{hillier98}
\bibinfo{author}{{Hillier} DJ} and  \bibinfo{author}{{Miller} DL} (\bibinfo{year}{1998}), \bibinfo{month}{Mar.}
\bibinfo{title}{{The Treatment of Non-LTE Line Blanketing in Spherically Expanding Outflows}}.
\bibinfo{journal}{{\em ApJ}} \bibinfo{volume}{496} (\bibinfo{number}{1}): \bibinfo{pages}{407--427}. \bibinfo{doi}{\doi{10.1086/305350}}.

\bibtype{Article}%
\bibitem[{Hubeny}(1988)]{hubeny88}
\bibinfo{author}{{Hubeny} I} (\bibinfo{year}{1988}), \bibinfo{month}{Dec.}
\bibinfo{title}{{A computer program for calculating non-LTE model stellar atmospheres}}.
\bibinfo{journal}{{\em Computer Physics Communications}} \bibinfo{volume}{52} (\bibinfo{number}{1}): \bibinfo{pages}{103--132}. \bibinfo{doi}{\doi{10.1016/0010-4655(88)90177-4}}.

\bibtype{Article}%
\bibitem[{Hunter} et al.(2008)]{Hunter08}
\bibinfo{author}{{Hunter} I}, \bibinfo{author}{{Brott} I}, \bibinfo{author}{{Lennon} DJ}, \bibinfo{author}{{Langer} N}, \bibinfo{author}{{Dufton} PL}, \bibinfo{author}{{Trundle} C}, \bibinfo{author}{{Smartt} SJ}, \bibinfo{author}{{de Koter} A}, \bibinfo{author}{{Evans} CJ} and  \bibinfo{author}{{Ryans} RSI} (\bibinfo{year}{2008}), \bibinfo{month}{Mar.}
\bibinfo{title}{{The VLT FLAMES Survey of Massive Stars: Rotation and Nitrogen Enrichment as the Key to Understanding Massive Star Evolution}}.
\bibinfo{journal}{{\em ApJ}} \bibinfo{volume}{676} (\bibinfo{number}{1}): \bibinfo{pages}{L29}. \bibinfo{doi}{\doi{10.1086/587436}}.
\eprint{0711.2267}.

\bibtype{Book}%
\bibitem[{Kopal}(1959)]{kopal59}
\bibinfo{author}{{Kopal} Z} (\bibinfo{year}{1959}).
\bibinfo{title}{{Close binary systems}}.

\bibtype{Article}%
\bibitem[{Lindegren} et al.(2021)]{lindegren21}
\bibinfo{author}{{Lindegren} L}, \bibinfo{author}{{Bastian} U}, \bibinfo{author}{{Biermann} M}, \bibinfo{author}{{Bombrun} A}, \bibinfo{author}{{de Torres} A}, \bibinfo{author}{{Gerlach} E}, \bibinfo{author}{{Geyer} R}, \bibinfo{author}{{Hern{\'a}ndez} J}, \bibinfo{author}{{Hilger} T}, \bibinfo{author}{{Hobbs} D}, \bibinfo{author}{{Klioner} SA}, \bibinfo{author}{{Lammers} U}, \bibinfo{author}{{McMillan} PJ}, \bibinfo{author}{{Ramos-Lerate} M}, \bibinfo{author}{{Steidelm{\"u}ller} H}, \bibinfo{author}{{Stephenson} CA} and  \bibinfo{author}{{van Leeuwen} F} (\bibinfo{year}{2021}), \bibinfo{month}{May}.
\bibinfo{title}{{Gaia Early Data Release 3. Parallax bias versus magnitude, colour, and position}}.
\bibinfo{journal}{{\em A\&A}} \bibinfo{volume}{649}, \bibinfo{eid}{A4}. \bibinfo{doi}{\doi{10.1051/0004-6361/202039653}}.
\eprint{2012.01742}.

\bibtype{Article}%
\bibitem[{Maeder} and {Meynet}(2000)]{maeder00}
\bibinfo{author}{{Maeder} A} and  \bibinfo{author}{{Meynet} G} (\bibinfo{year}{2000}), \bibinfo{month}{Sep.}
\bibinfo{title}{{Stellar evolution with rotation. VI. The Eddington and Omega -limits, the rotational mass loss for OB and LBV stars}}.
\bibinfo{journal}{{\em A\&A}} \bibinfo{volume}{361}: \bibinfo{pages}{159--166}. \bibinfo{doi}{\doi{10.48550/arXiv.astro-ph/0006405}}.
\eprint{astro-ph/0006405}.

\bibtype{Article}%
\bibitem[{Mahy} et al.(2015)]{mahy15}
\bibinfo{author}{{Mahy} L}, \bibinfo{author}{{Rauw} G}, \bibinfo{author}{{De Becker} M}, \bibinfo{author}{{Eenens} P} and  \bibinfo{author}{{Flores} CA} (\bibinfo{year}{2015}), \bibinfo{month}{May}.
\bibinfo{title}{{A spectroscopic investigation of the O-type star population in four Cygnus OB associations. II. Determination of the fundamental parameters}}.
\bibinfo{journal}{{\em A\&A}} \bibinfo{volume}{577}, \bibinfo{eid}{A23}. \bibinfo{doi}{\doi{10.1051/0004-6361/201321985}}.
\eprint{1504.03107}.

\bibtype{Article}%
\bibitem[{Mahy} et al.(2020)]{mahy20}
\bibinfo{author}{{Mahy} L}, \bibinfo{author}{{Almeida} LA}, \bibinfo{author}{{Sana} H}, \bibinfo{author}{{Clark} JS}, \bibinfo{author}{{de Koter} A}, \bibinfo{author}{{de Mink} SE}, \bibinfo{author}{{Evans} CJ}, \bibinfo{author}{{Grin} NJ}, \bibinfo{author}{{Langer} N}, \bibinfo{author}{{Moffat} AFJ}, \bibinfo{author}{{Schneider} FRN}, \bibinfo{author}{{Shenar} T} and  \bibinfo{author}{{Tramper} F} (\bibinfo{year}{2020}), \bibinfo{month}{Feb.}
\bibinfo{title}{{The Tarantula Massive Binary Monitoring. IV. Double-lined photometric binaries}}.
\bibinfo{journal}{{\em A\&A}} \bibinfo{volume}{634}, \bibinfo{eid}{A119}. \bibinfo{doi}{\doi{10.1051/0004-6361/201936152}}.
\eprint{1912.06853}.

\bibtype{Article}%
\bibitem[{Ma{\'\i}z Apell{\'a}niz}(2022)]{maizapellaniz22}
\bibinfo{author}{{Ma{\'\i}z Apell{\'a}niz} J} (\bibinfo{year}{2022}), \bibinfo{month}{Jan.}
\bibinfo{title}{{An estimation of the Gaia EDR3 parallax bias from stellar clusters and Magellanic Clouds data}}.
\bibinfo{journal}{{\em A\&A}} \bibinfo{volume}{657}, \bibinfo{eid}{A130}. \bibinfo{doi}{\doi{10.1051/0004-6361/202142365}}.
\eprint{2110.01475}.

\bibtype{Article}%
\bibitem[{Ma{\'\i}z Apell{\'a}niz} and {Barb{\'a}}(2018)]{maiz18}
\bibinfo{author}{{Ma{\'\i}z Apell{\'a}niz} J} and  \bibinfo{author}{{Barb{\'a}} RH} (\bibinfo{year}{2018}), \bibinfo{month}{May}.
\bibinfo{title}{{Optical-NIR dust extinction towards Galactic O stars}}.
\bibinfo{journal}{{\em A\&A}} \bibinfo{volume}{613}, \bibinfo{eid}{A9}. \bibinfo{doi}{\doi{10.1051/0004-6361/201732050}}.
\eprint{1712.09228}.

\bibtype{Article}%
\bibitem[{Mamajek} et al.(2015)]{mamajek15}
\bibinfo{author}{{Mamajek} EE}, \bibinfo{author}{{Torres} G}, \bibinfo{author}{{Prsa} A}, \bibinfo{author}{{Harmanec} P}, \bibinfo{author}{{Asplund} M}, \bibinfo{author}{{Bennett} PD}, \bibinfo{author}{{Capitaine} N}, \bibinfo{author}{{Christensen-Dalsgaard} J}, \bibinfo{author}{{Depagne} E}, \bibinfo{author}{{Folkner} WM}, \bibinfo{author}{{Haberreiter} M}, \bibinfo{author}{{Hekker} S}, \bibinfo{author}{{Hilton} JL}, \bibinfo{author}{{Kostov} V}, \bibinfo{author}{{Kurtz} DW}, \bibinfo{author}{{Laskar} J}, \bibinfo{author}{{Mason} BD}, \bibinfo{author}{{Milone} EF}, \bibinfo{author}{{Montgomery} MM}, \bibinfo{author}{{Richards} MT}, \bibinfo{author}{{Schou} J} and  \bibinfo{author}{{Stewart} SG} (\bibinfo{year}{2015}), \bibinfo{month}{Oct.}
\bibinfo{title}{{IAU 2015 Resolution B2 on Recommended Zero Points for the Absolute and Apparent Bolometric Magnitude Scales}}.
\bibinfo{journal}{{\em arXiv e-prints}} , \bibinfo{eid}{arXiv:1510.06262}\bibinfo{doi}{\doi{10.48550/arXiv.1510.06262}}.
\eprint{1510.06262}.

\bibtype{Article}%
\bibitem[{Marchant} and {Bodensteiner}(2023)]{marchant23}
\bibinfo{author}{{Marchant} P} and  \bibinfo{author}{{Bodensteiner} J} (\bibinfo{year}{2023}), \bibinfo{month}{Nov.}
\bibinfo{title}{{The Evolution of Massive Binary Stars}}.
\bibinfo{journal}{{\em arXiv e-prints}} , \bibinfo{eid}{arXiv:2311.01865}\bibinfo{doi}{\doi{10.48550/arXiv.2311.01865}}.
\eprint{2311.01865}.

\bibtype{Article}%
\bibitem[{Martig} et al.(2016)]{martig16}
\bibinfo{author}{{Martig} M}, \bibinfo{author}{{Fouesneau} M}, \bibinfo{author}{{Rix} HW}, \bibinfo{author}{{Ness} M}, \bibinfo{author}{{M{\'e}sz{\'a}ros} S}, \bibinfo{author}{{Garc{\'\i}a-Hern{\'a}ndez} DA}, \bibinfo{author}{{Pinsonneault} M}, \bibinfo{author}{{Serenelli} A}, \bibinfo{author}{{Silva Aguirre} V} and  \bibinfo{author}{{Zamora} O} (\bibinfo{year}{2016}), \bibinfo{month}{Mar.}
\bibinfo{title}{{Red giant masses and ages derived from carbon and nitrogen abundances}}.
\bibinfo{journal}{{\em MNRAS}} \bibinfo{volume}{456} (\bibinfo{number}{4}): \bibinfo{pages}{3655--3670}. \bibinfo{doi}{\doi{10.1093/mnras/stv2830}}.
\eprint{1511.08203}.

\bibtype{Article}%
\bibitem[{Martins} and {Palacios}(2013)]{martins13}
\bibinfo{author}{{Martins} F} and  \bibinfo{author}{{Palacios} A} (\bibinfo{year}{2013}), \bibinfo{month}{Dec.}
\bibinfo{title}{{A comparison of evolutionary tracks for single Galactic massive stars}}.
\bibinfo{journal}{{\em A\&A}} \bibinfo{volume}{560}, \bibinfo{eid}{A16}. \bibinfo{doi}{\doi{10.1051/0004-6361/201322480}}.
\eprint{1310.7218}.

\bibtype{Article}%
\bibitem[{Martins} et al.(2012)]{martins12}
\bibinfo{author}{{Martins} F}, \bibinfo{author}{{Mahy} L}, \bibinfo{author}{{Hillier} DJ} and  \bibinfo{author}{{Rauw} G} (\bibinfo{year}{2012}), \bibinfo{month}{Feb.}
\bibinfo{title}{{A quantitative study of O stars in NGC 2244 and the Monoceros OB2 association}}.
\bibinfo{journal}{{\em A\&A}} \bibinfo{volume}{538}, \bibinfo{eid}{A39}. \bibinfo{doi}{\doi{10.1051/0004-6361/201117458}}.
\eprint{1110.4509}.

\bibtype{Article}%
\bibitem[{Michelson} and {Pease}(1921)]{michelson20}
\bibinfo{author}{{Michelson} AA} and  \bibinfo{author}{{Pease} FG} (\bibinfo{year}{1921}), \bibinfo{month}{May}.
\bibinfo{title}{{Measurement of the Diameter of {\ensuremath{\alpha}} Orionis with the Interferometer.}}
\bibinfo{journal}{{\em ApJ}} \bibinfo{volume}{53}: \bibinfo{pages}{249--259}. \bibinfo{doi}{\doi{10.1086/142603}}.

\bibtype{Article}%
\bibitem[{Moe} and {Di Stefano}(2017)]{moe17}
\bibinfo{author}{{Moe} M} and  \bibinfo{author}{{Di Stefano} R} (\bibinfo{year}{2017}), \bibinfo{month}{Jun.}
\bibinfo{title}{{Mind Your Ps and Qs: The Interrelation between Period (P) and Mass-ratio (Q) Distributions of Binary Stars}}.
\bibinfo{journal}{{\em ApJS}} \bibinfo{volume}{230} (\bibinfo{number}{2}), \bibinfo{eid}{15}. \bibinfo{doi}{\doi{10.3847/1538-4365/aa6fb6}}.
\eprint{1606.05347}.

\bibtype{Article}%
\bibitem[{Ness} et al.(2016)]{ness16}
\bibinfo{author}{{Ness} M}, \bibinfo{author}{{Hogg} DW}, \bibinfo{author}{{Rix} HW}, \bibinfo{author}{{Martig} M}, \bibinfo{author}{{Pinsonneault} MH} and  \bibinfo{author}{{Ho} AYQ} (\bibinfo{year}{2016}), \bibinfo{month}{Jun.}
\bibinfo{title}{{Spectroscopic Determination of Masses (and Implied Ages) for Red Giants}}.
\bibinfo{journal}{{\em ApJ}} \bibinfo{volume}{823} (\bibinfo{number}{2}), \bibinfo{eid}{114}. \bibinfo{doi}{\doi{10.3847/0004-637X/823/2/114}}.
\eprint{1511.08204}.

\bibtype{Article}%
\bibitem[{Puls} et al.(2005)]{puls05}
\bibinfo{author}{{Puls} J}, \bibinfo{author}{{Urbaneja} MA}, \bibinfo{author}{{Venero} R}, \bibinfo{author}{{Repolust} T}, \bibinfo{author}{{Springmann} U}, \bibinfo{author}{{Jokuthy} A} and  \bibinfo{author}{{Mokiem} MR} (\bibinfo{year}{2005}), \bibinfo{month}{May}.
\bibinfo{title}{{Atmospheric NLTE-models for the spectroscopic analysis of blue stars with winds. II. Line-blanketed models}}.
\bibinfo{journal}{{\em A\&A}} \bibinfo{volume}{435} (\bibinfo{number}{2}): \bibinfo{pages}{669--698}. \bibinfo{doi}{\doi{10.1051/0004-6361:20042365}}.
\eprint{astro-ph/0411398}.

\bibtype{Article}%
\bibitem[{Royer} et al.(2024)]{royer24}
\bibinfo{author}{{Royer} P}, \bibinfo{author}{{Merle} T}, \bibinfo{author}{{Dsilva} K}, \bibinfo{author}{{Sekaran} S}, \bibinfo{author}{{Van Winckel} H}, \bibinfo{author}{{Fr{\'e}mat} Y}, \bibinfo{author}{{Van der Swaelmen} M}, \bibinfo{author}{{Gebruers} S}, \bibinfo{author}{{Tkachenko} A}, \bibinfo{author}{{Laverick} M}, \bibinfo{author}{{Dirickx} M}, \bibinfo{author}{{Raskin} G}, \bibinfo{author}{{Hensberge} H}, \bibinfo{author}{{Abdul-Masih} M}, \bibinfo{author}{{Acke} B}, \bibinfo{author}{{Alonso} ML}, \bibinfo{author}{{Bandhu Mahato} S}, \bibinfo{author}{{Beck} PG}, \bibinfo{author}{{Behara} N}, \bibinfo{author}{{Bloemen} S}, \bibinfo{author}{{Buysschaert} B}, \bibinfo{author}{{Cox} N}, \bibinfo{author}{{Debosscher} J}, \bibinfo{author}{{De Cat} P}, \bibinfo{author}{{Degroote} P}, \bibinfo{author}{{De Nutte} R}, \bibinfo{author}{{De Smedt} K}, \bibinfo{author}{{de Vries} B}, \bibinfo{author}{{Dumortier} L}, \bibinfo{author}{{Escorza} A}, \bibinfo{author}{{Exter} K}, \bibinfo{author}{{Goriely} S},
  \bibinfo{author}{{Gorlova} N}, \bibinfo{author}{{Hillen} M}, \bibinfo{author}{{Homan} W}, \bibinfo{author}{{Jorissen} A}, \bibinfo{author}{{Kamath} D}, \bibinfo{author}{{Karjalainen} M}, \bibinfo{author}{{Karjalainen} R}, \bibinfo{author}{{Lampens} P}, \bibinfo{author}{{Lobel} A}, \bibinfo{author}{{Lombaert} R}, \bibinfo{author}{{Marcos-Arenal} P}, \bibinfo{author}{{Menu} J}, \bibinfo{author}{{Merges} F}, \bibinfo{author}{{Moravveji} E}, \bibinfo{author}{{Nemeth} P}, \bibinfo{author}{{Neyskens} P}, \bibinfo{author}{{Ostensen} R}, \bibinfo{author}{{P{\'a}pics} PI}, \bibinfo{author}{{Perez} J}, \bibinfo{author}{{Prins} S}, \bibinfo{author}{{Royer} S}, \bibinfo{author}{{Samadi-Ghadim} A}, \bibinfo{author}{{Sana} H}, \bibinfo{author}{{Sans Fuentes} A}, \bibinfo{author}{{Scaringi} S}, \bibinfo{author}{{Schmid} V}, \bibinfo{author}{{Siess} L}, \bibinfo{author}{{Siopis} C}, \bibinfo{author}{{Smolders} K}, \bibinfo{author}{{S{\'o}dor} {\'A}}, \bibinfo{author}{{Thoul} A}, \bibinfo{author}{{Triana} S},
  \bibinfo{author}{{Vandenbussche} B}, \bibinfo{author}{{Van de Sande} M}, \bibinfo{author}{{Van De Steene} G}, \bibinfo{author}{{Van Eck} S}, \bibinfo{author}{{van Hoof} PAM}, \bibinfo{author}{{Van Marle} AJ}, \bibinfo{author}{{Van Reeth} T}, \bibinfo{author}{{Vermeylen} L}, \bibinfo{author}{{Volpi} D}, \bibinfo{author}{{Vos} J} and  \bibinfo{author}{{Waelkens} C} (\bibinfo{year}{2024}), \bibinfo{month}{Jan.}
\bibinfo{title}{{MELCHIORS. The Mercator Library of High Resolution Stellar Spectroscopy}}.
\bibinfo{journal}{{\em A\&A}} \bibinfo{volume}{681}, \bibinfo{eid}{A107}. \bibinfo{doi}{\doi{10.1051/0004-6361/202346847}}.
\eprint{2311.02705}.

\bibtype{Article}%
\bibitem[{Sana} et al.(2012)]{sana12}
\bibinfo{author}{{Sana} H}, \bibinfo{author}{{de Mink} SE}, \bibinfo{author}{{de Koter} A}, \bibinfo{author}{{Langer} N}, \bibinfo{author}{{Evans} CJ}, \bibinfo{author}{{Gieles} M}, \bibinfo{author}{{Gosset} E}, \bibinfo{author}{{Izzard} RG}, \bibinfo{author}{{Le Bouquin} JB} and  \bibinfo{author}{{Schneider} FRN} (\bibinfo{year}{2012}), \bibinfo{month}{Jul.}
\bibinfo{title}{{Binary Interaction Dominates the Evolution of Massive Stars}}.
\bibinfo{journal}{{\em Science}} \bibinfo{volume}{337} (\bibinfo{number}{6093}): \bibinfo{pages}{444}. \bibinfo{doi}{\doi{10.1126/science.1223344}}.
\eprint{1207.6397}.

\bibtype{Article}%
\bibitem[{Schneider} et al.(2018)]{schneider18}
\bibinfo{author}{{Schneider} FRN}, \bibinfo{author}{{Ram{\'\i}rez-Agudelo} OH}, \bibinfo{author}{{Tramper} F}, \bibinfo{author}{{Bestenlehner} JM}, \bibinfo{author}{{Castro} N}, \bibinfo{author}{{Sana} H}, \bibinfo{author}{{Evans} CJ}, \bibinfo{author}{{Sab{\'\i}n-Sanjuli{\'a}n} C}, \bibinfo{author}{{Sim{\'o}n-D{\'\i}az} S}, \bibinfo{author}{{Langer} N}, \bibinfo{author}{{Fossati} L}, \bibinfo{author}{{Gr{\"a}fener} G}, \bibinfo{author}{{Crowther} PA}, \bibinfo{author}{{de Mink} SE}, \bibinfo{author}{{de Koter} A}, \bibinfo{author}{{Gieles} M}, \bibinfo{author}{{Herrero} A}, \bibinfo{author}{{Izzard} RG}, \bibinfo{author}{{Kalari} V}, \bibinfo{author}{{Klessen} RS}, \bibinfo{author}{{Lennon} DJ}, \bibinfo{author}{{Mahy} L}, \bibinfo{author}{{Ma{\'\i}z Apell{\'a}niz} J}, \bibinfo{author}{{Markova} N}, \bibinfo{author}{{van Loon} JT}, \bibinfo{author}{{Vink} JS} and  \bibinfo{author}{{Walborn} NR} (\bibinfo{year}{2018}), \bibinfo{month}{Oct.}
\bibinfo{title}{{The VLT-FLAMES Tarantula Survey. XXIX. Massive star formation in the local 30 Doradus starburst}}.
\bibinfo{journal}{{\em A\&A}} \bibinfo{volume}{618}, \bibinfo{eid}{A73}. \bibinfo{doi}{\doi{10.1051/0004-6361/201833433}}.
\eprint{1807.03821}.

\bibtype{Book}%
\bibitem[{Seidelmann}(1992)]{Seidelmann92}
\bibinfo{author}{{Seidelmann} PK} (\bibinfo{year}{1992}).
\bibinfo{title}{{Explanatory Supplement to the Astronomical Almanac}}.

\bibtype{Article}%
\bibitem[{Serenelli} et al.(2021)]{serenelli21}
\bibinfo{author}{{Serenelli} A}, \bibinfo{author}{{Weiss} A}, \bibinfo{author}{{Aerts} C}, \bibinfo{author}{{Angelou} GC}, \bibinfo{author}{{Baroch} D}, \bibinfo{author}{{Bastian} N}, \bibinfo{author}{{Beck} PG}, \bibinfo{author}{{Bergemann} M}, \bibinfo{author}{{Bestenlehner} JM}, \bibinfo{author}{{Czekala} I}, \bibinfo{author}{{Elias-Rosa} N}, \bibinfo{author}{{Escorza} A}, \bibinfo{author}{{Van Eylen} V}, \bibinfo{author}{{Feuillet} DK}, \bibinfo{author}{{Gandolfi} D}, \bibinfo{author}{{Gieles} M}, \bibinfo{author}{{Girardi} L}, \bibinfo{author}{{Lebreton} Y}, \bibinfo{author}{{Lodieu} N}, \bibinfo{author}{{Martig} M}, \bibinfo{author}{{Miller Bertolami} MM}, \bibinfo{author}{{Mombarg} JSG}, \bibinfo{author}{{Morales} JC}, \bibinfo{author}{{Moya} A}, \bibinfo{author}{{Nsamba} B}, \bibinfo{author}{{Pavlovski} K}, \bibinfo{author}{{Pedersen} MG}, \bibinfo{author}{{Ribas} I}, \bibinfo{author}{{Schneider} FRN}, \bibinfo{author}{{Silva Aguirre} V}, \bibinfo{author}{{Stassun} KG}, \bibinfo{author}{{Tolstoy} E},
  \bibinfo{author}{{Tremblay} PE} and  \bibinfo{author}{{Zwintz} K} (\bibinfo{year}{2021}), \bibinfo{month}{Dec.}
\bibinfo{title}{{Weighing stars from birth to death: mass determination methods across the HRD}}.
\bibinfo{journal}{{\em A\&A~Rev.}} \bibinfo{volume}{29} (\bibinfo{number}{1}), \bibinfo{eid}{4}. \bibinfo{doi}{\doi{10.1007/s00159-021-00132-9}}.
\eprint{2006.10868}.

\bibtype{Article}%
\bibitem[{Sim{\'o}n-D{\'\i}az} and {Herrero}(2007)]{simon-diaz07}
\bibinfo{author}{{Sim{\'o}n-D{\'\i}az} S} and  \bibinfo{author}{{Herrero} A} (\bibinfo{year}{2007}), \bibinfo{month}{Jun.}
\bibinfo{title}{{Fourier method of determining the rotational velocities in OB stars}}.
\bibinfo{journal}{{\em A\&A}} \bibinfo{volume}{468} (\bibinfo{number}{3}): \bibinfo{pages}{1063--1073}. \bibinfo{doi}{\doi{10.1051/0004-6361:20066060}}.
\eprint{astro-ph/0703216}.

\bibtype{Article}%
\bibitem[{Sim{\'o}n-D{\'\i}az} and {Herrero}(2014)]{simondiaz14}
\bibinfo{author}{{Sim{\'o}n-D{\'\i}az} S} and  \bibinfo{author}{{Herrero} A} (\bibinfo{year}{2014}), \bibinfo{month}{Feb.}
\bibinfo{title}{{The IACOB project. I. Rotational velocities in northern Galactic O- and early B-type stars revisited. The impact of other sources of line-broadening}}.
\bibinfo{journal}{{\em A\&A}} \bibinfo{volume}{562}, \bibinfo{eid}{A135}. \bibinfo{doi}{\doi{10.1051/0004-6361/201322758}}.
\eprint{1311.3360}.

\bibtype{Article}%
\bibitem[{Tkachenko} et al.(2020)]{tkachenko20}
\bibinfo{author}{{Tkachenko} A}, \bibinfo{author}{{Pavlovski} K}, \bibinfo{author}{{Johnston} C}, \bibinfo{author}{{Pedersen} MG}, \bibinfo{author}{{Michielsen} M}, \bibinfo{author}{{Bowman} DM}, \bibinfo{author}{{Southworth} J}, \bibinfo{author}{{Tsymbal} V} and  \bibinfo{author}{{Aerts} C} (\bibinfo{year}{2020}), \bibinfo{month}{May}.
\bibinfo{title}{{The mass discrepancy in intermediate- and high-mass eclipsing binaries: The need for higher convective core masses}}.
\bibinfo{journal}{{\em A\&A}} \bibinfo{volume}{637}, \bibinfo{eid}{A60}. \bibinfo{doi}{\doi{10.1051/0004-6361/202037452}}.
\eprint{2003.08982}.

\bibtype{Article}%
\bibitem[{Torres} et al.(2010)]{torres10}
\bibinfo{author}{{Torres} G}, \bibinfo{author}{{Andersen} J} and  \bibinfo{author}{{Gim{\'e}nez} A} (\bibinfo{year}{2010}), \bibinfo{month}{Feb.}
\bibinfo{title}{{Accurate masses and radii of normal stars: modern results and applications}}.
\bibinfo{journal}{{\em A\&A~Rev.}} \bibinfo{volume}{18} (\bibinfo{number}{1-2}): \bibinfo{pages}{67--126}. \bibinfo{doi}{\doi{10.1007/s00159-009-0025-1}}.
\eprint{0908.2624}.

\bibtype{Inproceedings}%
\bibitem[{van der Hucht}(1996)]{vanderhucht96}
\bibinfo{author}{{van der Hucht} KA} (\bibinfo{year}{1996}), \bibinfo{month}{Jan.}, \bibinfo{title}{{Past and present classification of hot massive stars}}, \bibinfo{editor}{{Vreux} JM}, \bibinfo{editor}{{Detal} A}, \bibinfo{editor}{{Fraipont-Caro} D}, \bibinfo{editor}{{Gosset} E} and  \bibinfo{editor}{{Rauw} G}, (Eds.), \bibinfo{booktitle}{Liege International Astrophysical Colloquia}, \bibinfo{series}{Liege International Astrophysical Colloquia}, \bibinfo{volume}{33}, pp.~\bibinfo{pages}{1}.

\bibtype{Article}%
\bibitem[{Willmer}(2018)]{willmer18}
\bibinfo{author}{{Willmer} CNA} (\bibinfo{year}{2018}), \bibinfo{month}{Jun.}
\bibinfo{title}{{The Absolute Magnitude of the Sun in Several Filters}}.
\bibinfo{journal}{{\em ApJS}} \bibinfo{volume}{236} (\bibinfo{number}{2}), \bibinfo{eid}{47}. \bibinfo{doi}{\doi{10.3847/1538-4365/aabfdf}}.
\eprint{1804.07788}.

\end{thebibliography*}

\end{document}